\documentclass[10pt,aps,twocolumn,prd,notitlepage,amssymb,amsmath,floatfix,nofootinbib,superscriptaddress]{revtex4-1}

\usepackage{epsfig}
\usepackage{bm}
\usepackage{amssymb}
\usepackage{amsmath}
\usepackage{color}
\usepackage{framed}
\usepackage{xcolor}
\usepackage[colorlinks,linkcolor=blue,anchorcolor=black,citecolor=blue]{hyperref}
\usepackage{multirow}

\usepackage[T1]{fontenc}
\usepackage[latin9]{inputenc}
\usepackage{amsmath}
\usepackage{graphicx}

\allowdisplaybreaks[4]

\mathchardef\mhyphen="2D

\makeatletter
\@ifundefined{textcolor}{}
{%
 \definecolor{BLACK}{gray}{0}
 \definecolor{WHITE}{gray}{1}
 \definecolor{RED}{rgb}{1,0,0}
 \definecolor{GREEN}{rgb}{0,1,0}
 \definecolor{BLUE}{rgb}{0,0,1}
 \definecolor{CYAN}{cmyk}{1,0,0,0}
 \definecolor{MAGENTA}{cmyk}{0,1,0,0}
 \definecolor{YELLOW}{cmyk}{0,0,1,0}
}

\makeatother
\usepackage{babel}

\begin{document}
\title{Global quark spin correlations in relativistic heavy ion collisions}
\author{Ji-peng Lv}
\email{jipenglv@mail.sdu.edu.cn}
\affiliation{Institute of Frontier and Interdisciplinary Science and Key Laboratory
of Particle Physics and Particle Irradiation (MOE), Shandong University, Qingdao, Shandong 266237, China}

\author{Zi-han Yu}
\email{zihan.yu@mail.sdu.edu.cn}
\affiliation{Institute of Frontier and Interdisciplinary Science and Key Laboratory
of Particle Physics and Particle Irradiation (MOE), Shandong University, Qingdao, Shandong 266237, China}

\author{Zuo-tang Liang}
\email{liang@sdu.edu.cn}
\affiliation{Institute of Frontier and Interdisciplinary Science and Key Laboratory
of Particle Physics and Particle Irradiation (MOE), Shandong University, Qingdao, Shandong 266237, China}

\author{Qun Wang}
\email{qunwang@ustc.edu.cn}
\affiliation{Department of Modern Physics, University of Science and Technology
of China, Hefei, Anhui 230026, China}
\affiliation{School of Mechanics and Physics, Anhui University of Science and Technology,
Huainan, Anhui 232001, China}

\author{Xin-Nian Wang}
\email{xnwang@lbl.gov}
\affiliation{Nuclear Science Division, MS 70R0319, Lawrence Berkeley National Laboratory,
Berkeley, California 94720, USA}
\begin{abstract}
The observation of the vector meson's global spin alignment by the
STAR Collaboration reveals that strong spin correlations may exist
for quarks and antiquarks in relativistic heavy-ion collisions 
in the normal direction of the reaction plane. 
We propose a systematic method to describe such correlations in the quark matter. 
The correlations can be classified as local and long range types. 
We show in particular that the effective quark spin correlations contain 
the genuine spin correlations originated directly from the dynamical process as well as those induced by averaging over other degrees of freedom. 
We also show that such correlations can be studied by measuring the vector meson's spin density matrix
and hyperon-hyperon and hyperon-anti-hyperon spin correlations. 
We present the relationships between these measurable quantities and spin correlations of quarks and antiquarks. 
\end{abstract}
\maketitle

\section{Introduction}

The global hyperon polarization has been observed first by the STAR
Collaboration at the Relativistic Heavy Ion Collider (RHIC)~\citep{STAR:2017ckg}
and later in a series of subsequent experiments~\citep{STAR:2018gyt,STAR:2020xbm,STAR:2021beb,ALICE:2019onw,HADES:2022enx}.
This confirms the theoretical predictions made almost two decades ago~\citep{Liang:2004ph,Liang:2004xn,Gao:2007bc}. 
The experimental results can be described quantitatively using phenomenological transport or hydrodynamical models~\citep{Betz:2007kg,Becattini:2007sr,Karpenko:2016jyx,Pang:2016igs,Li:2017slc,Xie:2017upb,Sun:2017xhx,Baznat:2017jfj,Shi:2017wpk,Xia:2018tes,Wei:2018zfb,Fu:2020oxj,Ryu:2021lnx,Fu:2021pok,Deng:2021miw,Becattini:2021iol,Wu:2022mkr},
that are reviewed e.g. in~\citep{Gao:2020lxh,Huang:2020dtn,Becattini:2021lfq,Florkowski:2018fap,Becattini:2020ngo,Gao:2020vbh,Becattini:2022zvf,Jian-Hua:2023cna,Liang:2019clf,Liang:2007ma,Wang:2017jpl,Huang:2020xyr,Liang:2022ekv,Dong:2024nxj}.
Such studies open a new avenue in studying properties of the quark gluon plasma (QGP) produced in heavy ion collisions and have attracted much attention in the field. 

Recently, the STAR Collaboration published their measurements on vector mesons' global spin alignment~\citep{STAR:2022fan}. 
Their results, together with other measurements~\citep{STAR:2008lcm,ALICE:2019aid,ALICE:2022byg,ALICE:2022dyy}
bring the field of spin physics in heavy-ion collisions to a new climax~\citep{Wang:2023fvy,Chen:2023hnb,talks,Li-Juan:2023bws,Xin-Li:2023gwh,Yang:2017sdk,Sheng:2019kmk,Sheng:2020ghv,Sheng:2022wsy,Sheng:2022ffb,Xia:2020tyd,Wei:2023pdf,Kumar:2023ghs,DeMoura:2023jzz,Fu:2023qht,Sheng:2023urn,Fang:2023bbw,Dong:2023cng,Kumar:2023ojl}.
The STAR results~\citep{STAR:2022fan}, on one hand, show that global
polarization effects also exist for vector mesons, on the other hand,
they seem to be inconsistent with the magnitude of hyperon polarization if quark spin correlations due to strong fields are neglected.
As was shown in the original papers on the global polarization effect~\citep{Liang:2004ph,Liang:2004xn,Gao:2007bc}, 
if we take quark polarization as a constant and neglect fluctuations and correlations etc., 
the hyperon polarization should be equal to that of the quark but the vector meson's spin alignment should be proportional to the quark polarization squared.
In this case, since the observed hyperon polarization is only a few
percent~\citep{STAR:2017ckg,STAR:2018gyt,STAR:2020xbm,STAR:2021beb,ALICE:2019onw,HADES:2022enx},
the vector meson's spin alignment should be much smaller than those
observed in the STAR experiment~\citep{STAR:2022fan}. 
Hence, the data clearly reveal that there is strong spin correlation between
the quark and anti-quark that combine into the vector meson~\citep{Wang:2023fvy,Chen:2023hnb,talks,Li-Juan:2023bws,Xin-Li:2023gwh,Yang:2017sdk,Sheng:2019kmk,Sheng:2020ghv,Sheng:2022wsy,Sheng:2022ffb,Xia:2020tyd,Wei:2023pdf,Kumar:2023ghs,DeMoura:2023jzz,Fu:2023qht,Sheng:2023urn,Fang:2023bbw,Dong:2023cng,Kumar:2023ojl}.
In this sense, this provides the first opportunity to study the spin correlations at the quark level in high energy heavy-ion collisions. 

It is clear that the spin correlation of quarks and antiquarks is an important property of QGP. It may contain important information
on strong interaction and provide new clue to color confinement in quantum chromodynamics (QCD). So it is crucial to present a unified
and description of spin correlations in quark matter and make connection with experimental observables.

The spin correlation in a system of spin-1/2 particles have been defined
differently in text books. However, to study spin properties of QGP,
it is convenient to give a definition in a way that one can study
the correlation order by order. Such a definition will facilitate
the study and reveal the underlying dynamics. The connection between the
spin correlation and experimental observables depend on hadronization mechanism
and thus may be model dependent.

The purpose of this paper is to propose a systematic description of
spin correlation of quarks and anti-quarks in heavy-ion collisions.
We show that the spin correlation can be decomposed into the genuine
and induced one and also into the local and long range one. We propose
that the spin correlation at the quark level can be extracted from
the vector meson's diagonal and off-diagonal elements of its spin
density matrix together with the hyperon-anti-hyperon spin correlation.
We present the relationships between the spin correlation at the quark
level and measurable quantities in a simple quark recombination model.

The rest of the paper is organized as follows. In Sec.~\ref{sec:qc},
we propose the systematic way to describe quark spin correlation in
the quark matter system and discuss its properties. In Sec.~\ref{sec:rhoV},
we present the results of the vector meson's spin alignment and off-diagonal
elements of the spin density matrix as functions of the quark spin
polarization and correlation. In Sec.~\ref{sec:Hpol}, we present
the results for the hyperon polarization and hyperon-hyperon or hyperon-anti-hyperon
spin correlation. We also present numerical estimates of spin correlation
parameters by fitting the existing data in Sec.~\ref{sec:numerical}.
Finally, a short summary and an outlook are given in Sec.~\ref{sec:summary}.

\section{Quark spin correlations in quark matter system \label{sec:qc}}

We consider a quark matter system such as QGP consisting of quarks and anti-quarks. 
The spin properties of the system are described by the spin density matrix. 
For a single particle, we study the spin polarization while for two or more particles 
we can study not only the spin polarization but also the spin correlation. 

\subsection{The spin density matrix}\label{sec:qrho}

In a spin-1/2 particle system, the spin state of a particle is described by the spin density matrix 
that can be expanded in terms of the complete set of the $2\times 2$ hermitian matrices \{$\mathbb{I}$, $\hat\sigma_i$\}, i.e., 
\begin{align}
&\hat\rho^{(q)}=\frac{1}{2}(\mathbb{I}+P_{qi}\hat\sigma_i), \label{eq:rho1/2}
\end{align} 
where $P_{qi}=\langle\hat\sigma_i\rangle={\rm Tr}[\hat\rho^{(q)}\hat{\sigma_i}]$ with 
$i=x,y,z$ is the $i$-th component of the quark polarization vector $\boldsymbol{P}_{q}=(P_{qx},P_{qy},P_{qz})$, 
$\hat\sigma_{i}$ denotes Pauli matrices and ${\rm Tr}\hat\rho^{(q)}=1$ is normalized to one.
The symbol $\mathbb{I}$ denotes the unit $2\times 2$ matrix, 
and in the following of this paper, we simply write it as $1$. 
Also, we use the convention that a sum over repeated indices is implicit through out the paper.

For two particle state in the system, we denote the spin density matrix by $\hat\rho^{(12)}$. 
Conventionally, one expends $\hat\rho^{(12)}$ in terms of the complete set of hermitian matrices 
$\{1\otimes 1, \hat\sigma_{1i}\otimes 1, 1\otimes\hat\sigma_ {2j}, \hat\sigma_{1i}\otimes\hat\sigma_{2j}\}$,  i.e., 
\begin{equation}
\hat{\rho}^{(12)}=\frac{1}{4}\left[1+P_{1i}\hat\sigma_{1i}+P_{2j}\hat\sigma_{2j}+t_{ij}^{(12)}\hat\sigma_{1i}\otimes\hat\sigma_{2j}\right], \label{eq:trho12}
\end{equation}
where $t_{ij}^{(12)}$ is called the spin correlation of particles 1 and 2.
Here as well as in the following of this paper, we take the following scheme of notations: 
the superscript of the spin density matrix $\hat\rho$ or the spin correlation $t$ denotes the type of particle 
where for hadrons we simply use the symbol while for quarks or anti-quarks we put it in a bracket; 
the subscript of them denotes the indices of matrix elements or spatial components such as $mm'$ or $ij$.  
For polarization vectors, we simply use double subscripts to specify particle type and spatial component respectively.    

There is however a shortcoming in the definition of the spin correlation
through $t_{ij}^{(12)}$ in Eq.~(\ref{eq:trho12}). In case of no spin correlation
between particles 1 and 2, we should have $\hat{\rho}^{(12)}=\hat{\rho}^{(1)}\otimes\hat{\rho}^{(2)}$
and then $t_{ij}^{(12)}=P_{1i}P_{2j}$ that is non-vanishing.
The situation is the same for spin correlations of three or more particles
if they are defined in a similar way. This is in particular inconvenient
if we study the spin correlations order by order.

To overcome such a shortcoming, we propose to expend $\hat{\rho}^{(12)}$ in the following way, 
\begin{align}
\hat\rho^{(12)}=\hat\rho^{(1)}\otimes\hat\rho^{(2)}+\frac{1}{2^2}c_{ij}^{(12)} \hat\sigma_{1i}\otimes\hat\sigma_{2j}, \label{eq:rho12}
\end{align} 
where the spin correlation is described by $c_{ij}^{(12)}$. It is
clear that $c_{ij}^{(12)}=0$ if there is no spin correlation between
particles 1 and 2. In the same way, we expand the spin density matrix
for a system of three or four particles as 
\begin{widetext}
\begin{align}
\hat\rho^{(123)}=& \hat\rho^{(1)} \otimes \hat\rho^{(2)} \otimes \hat\rho^{(3)}  +\frac{1}{2^3} c_{ijk}^{(123)} \hat\sigma_{1i}\otimes\hat\sigma_{2j}\otimes\hat\sigma_{3k} \nonumber\\
& + \frac{1}{2^2} \Bigl[ c_{ij}^{(12)} \hat\sigma_{1i}\otimes\hat\sigma_{2j}\otimes \hat\rho^{(3)}  
+ c_{jk}^{(23)} \hat\rho^{(1)}  \otimes\hat\sigma_{2j}\otimes\hat\sigma_{3k}
+c_{ik}^{(13)} \hat\sigma_{1i}\otimes\hat\rho^{(2)} \otimes\hat\sigma_{3k}  \Bigr], \label{eq:rho123} \\
\hat\rho^{(1234)}=& \hat\rho^{(1)} \otimes \hat\rho^{(2)} \otimes \hat\rho^{(3)} \otimes \hat\rho^{(4)}+ \frac{1}{2^4} c_{ijkl}^{(1234)} \hat\sigma_{1i}\otimes\hat\sigma_{2j}\otimes\hat\sigma_{3k}\otimes\hat\sigma_{4l} \nonumber\\
&+ \frac{1}{2^2}\Bigl[c_{ij}^{(12)} \hat\sigma_{1i}\otimes\hat\sigma_{2j}\otimes \hat\rho^{(3)} \otimes \hat\rho^{(4)}
+c_{kl}^{(34)} \hat\rho^{(1)} \otimes \hat\rho^{(2)} \otimes \hat\sigma_{3k}\otimes\hat\sigma_{4l} +c_{ik}^{(13)} \hat\sigma_{1i}\otimes\hat\rho^{(2)}\otimes\hat\sigma_{3k}\otimes\hat\rho^{(4)} \nonumber\\
&~~~~~~~ +c_{jl}^{(24)} \hat\rho^{(1)}\otimes\hat\sigma_{2j}\otimes\hat\rho^{(3)}\otimes \hat\sigma_{4l} 
+c_{il}^{(14)} \hat\sigma_{1i}\otimes\hat\rho^{(2)}\otimes\hat\rho^{(3)} \otimes\hat\sigma_{4l}
+c_{jk}^{(23)} \hat\rho^{(1)}\otimes\hat\sigma_{2j}\otimes \hat\sigma_{3k}\otimes\hat\rho^{(4)} \Bigr] \nonumber\\
&+\frac{1}{2^3}\Bigl[ c_{ijk}^{(123)} \hat\sigma_{1i}\otimes\hat\sigma_{2j}\otimes\hat\sigma_{3k}\otimes\hat\rho^{(4)} 
+c_{ijl}^{(124)} \hat\sigma_{1i}\otimes\hat\sigma_{2j}\otimes\hat\rho^{(3)} \otimes\hat\sigma_{4l} +c_{ikl}^{(134)} \hat\sigma_{1i}\otimes\hat\rho^{(2)}\otimes\hat\sigma_{3k}\otimes\hat\sigma_{4l} \nonumber\\
&~~~~~~~ +c_{jkl}^{(234)} \hat\rho^{(1)} \otimes\hat\sigma_{2j}\otimes\hat\sigma_{3k}\otimes\hat\sigma_{4l} \Bigr] .  \label{eq:rho1234}
\end{align}

The polarizations and spin correlations can be extracted by taking
expectation values of a direct product of Pauli matrices on spin density
matrices. The results are 
\begin{align}
P_{1i}=&\langle\hat\sigma_{1i}\rangle, \\
c^{(12)}_{ij}=&\langle\hat\sigma_{1i}\hat\sigma_{2j}\rangle-P_{1i}P_{2j}, \\
c^{(123)}_{ijk}=&\langle\hat\sigma_{1i}\hat\sigma_{2j}\hat\sigma_{3k}\rangle-c^{(12)}_{ij}P_{3k}-c^{(23)}_{jk}P_{1i}-c^{(13)}_{ik}P_{2j}-P_{1i}P_{2j}P_{3k}, \\
c^{(1234)}_{ijkl}=&\langle\hat\sigma_{1i}\hat\sigma_{2j}\hat\sigma_{3k}\hat\sigma_{4l}\rangle
-c^{(123)}_{ijk}P_{4l}-c^{(124)}_{ijl}P_{3k}-c^{(134)}_{ikl}P_{2j}-c^{(234)}_{jkl}P_{1i}
-c^{(12)}_{ij}P_{3k}P_{4l}-c^{(13)}_{ik}P_{2j}P_{4l}\nonumber\\ 
&-c^{(14)}_{il}P_{2j}P_{3k}-c^{(23)}_{jk}P_{1i}P_{4l}
-c^{(24)}_{jl}P_{1i}P_{3k}-c^{(34)}_{kl}P_{1i}P_{2j}-P_{1i}P_{2j}P_{3k}P_{4l}.
\end{align}

For a four-particle system, according to Eq.~(\ref{eq:rho1234}),
if the system do not have any spin correlations, i.e., the spin density
matrix of the system is the direct product of spin density matrices
of single particle, we have 
\begin{align}
 & c_{ij}^{(12)}=c_{jk}^{(23)}=c_{ik}^{(13)}=c_{kl}^{(34)}=c_{jl}^{(24)}=c_{il}^{(14)}=0,\nonumber \\
 & c_{ijk}^{(123)}=c_{jkl}^{(234)}=c_{ijl}^{(124)}=c_{ikl}^{(134)}=0,\nonumber \\
 & c_{ijkl}^{(1234)}=0.
\end{align}
If there are only two-particle spin correlations, we have 
\begin{equation}
c_{ijk}^{(123)}=c_{jkl}^{(234)}=c_{ijl}^{(124)}=c_{ikl}^{(134)}=c_{ijkl}^{(1234)}=0.
\end{equation}
If there are only two-particle and three-particle spin correlations, we
have $c_{ijkl}^{(1234)}=0$. In this way, we can include spin correlations order by order.

We note that if we define the spin correlation for two spin-$1/2$ particles $h_1h_2$ in the conventional way, i.e, 
\begin{equation}
c_{nn}=\frac{f_{++}+f_{--}-f_{+-}-f_{-+}}{f_{++}+f_{--}+f_{+-}+f_{-+}},\label{eq:cnndef}
\end{equation}
where $n$ stands for the spin quantization direction $\hat{\boldsymbol{n}}$,
$f_{m_1m_2}=\left\langle m_1m_2\right|\hat{\rho}^{(12)}\left|m_1m_2\right\rangle $
(with $m_1,m_2=\pm$ denoting spin states) is the fraction of the particle pair in the spin state $|m_1m_2\rangle $. 
We then obtain the relationship between $c_{nn}$ and $c_{ij}^{(12)}$ defined above as 
\begin{equation}
c_{nn}=c_{nn}^{(12)}+P_{1n}P_{2n}.
\end{equation}

\subsection{With other degrees of freedom \label{sec:qcalpha}}

We suppose particles in the system have other degrees of freedom that are denoted in general by $\alpha$. 
We consider here a very simple case that the polarization and spin correlations have $\alpha$-dependence
so that spin density matrices are given by 
\begin{align}
&\hat\rho^{(q)}(\alpha)=\frac{1}{2} \bigl[1+P_{qi}(\alpha) \hat\sigma_i\bigr], \label{eq:rho1/2alpha} \\
&\hat\rho^{(12)}(\alpha_1,\alpha_2)=\hat\rho^{(1)}(\alpha_1)\otimes \hat\rho^{(2)}(\alpha_2)
+\frac{1}{2^2} c_{ij}^{(12)}(\alpha_1,\alpha_2) \hat\sigma_{1i}\otimes\hat\sigma_{2j} . \label{eq:rho12alpha}
\end{align} 

Now suppose we have a system (12) composed of 1 and 2. 
We assume that the system is at the state $|\alpha_{12}\rangle$ so the probability to find particle 1 and 2 
at $\alpha_1$ and $\alpha_2$ respectively is determined by 
the amplitude $\left\langle \alpha_{1},\alpha_{2}\right.\left|\alpha_{12}\right\rangle $.
We obtain the effective spin density matrix for the system (12) at $\alpha_{12}$ as
\begin{align}
\hat{\bar\rho}^{(12)}(\alpha_{12})&=\langle\alpha_{12}|\hat\rho^{(12)}|\alpha_{12}\rangle 
= \hat{\bar\rho}^{(1)}(\alpha_{12})\otimes \hat{\bar\rho}^{(2)}(\alpha_{12})
+\frac{1}{2^2}  \bar c_{ij}^{(12)}(\alpha_{12})\hat\sigma_{1i}\otimes\hat\sigma_{2j} , \label{eq:rho12b}
\end{align} 
where the average of $\hat\rho^{(1)}(\alpha_{1})$ and that of $P_{1i}(\alpha_1)$ weighted by 
the wave function $\langle\alpha_{1},\alpha_{2}|\alpha_{12}\rangle$ squared are given by
\begin{align}
&\hat{\bar\rho}^{(1)}(\alpha_{12})= \langle \hat\rho^{(1)} (\alpha_{1})\rangle 
\equiv  \sum_{\alpha_1\alpha_2} |\langle\alpha_{1},\alpha_{2}|\alpha_{12}\rangle|^2 \hat\rho^{(1)} (\alpha_{1}) 
=\frac{1}{2} [1+\bar P_{1i} (\alpha_{12}) \hat\sigma_{1i}], \label{eq:brho1}\\
&\bar P_{1i} (\alpha_{12}) =\langle P_{1i} (\alpha_{1})\rangle 
\equiv \sum_{\alpha_1\alpha_2}|\langle\alpha_{1},\alpha_{2}|\alpha_{12}\rangle|^2 P_{1i} (\alpha_1).\label{eq:bP1i} 
\end{align}
Here as well as in the following of this paper we use $\langle\cdots\rangle$ to denote such an average on the state of the system.

In the case of $\bar c_{ij}^{(12)}(\alpha_{12})$, we have, 
\begin{align}
\bar c_{ij}^{(12)} (\alpha_{12}) 
&=\langle  c_{ij}^{(12)}(\alpha_1,\alpha_2) + P_{1i} (\alpha_1) P_{2j} (\alpha_2) \rangle - \bar P_{1i} (\alpha_{12}) \bar P_{2j} (\alpha_{12}). \label{eq:bc12}
\end{align}
We see that $\bar c_{ij}^{(12)} (\alpha_{12}) $ is not simply the average of $c_{ij}^{(12)}(\alpha_1,\alpha_2)$ 
weighted by the wave function $\langle\alpha_{1},\alpha_{2}|\alpha_{12}\rangle$ squared. 
In particular, in the case of $c_{ij}^{(12)}(\alpha_{1},\alpha_{2})=0$, we have
\begin{equation}
\bar{c}_{ij}^{(12;0)}(\alpha_{12})=\left\langle P_{1i}(\alpha_{1})P_{2j}(\alpha_{2})\right\rangle -\left\langle P_{1i}(\alpha_{1})\right\rangle \left\langle P_{2j}(\alpha_{2})\right\rangle. \label{eq:bc120}
\end{equation}

We see clearly that $\bar c_{ij}^{(12;0)} (\alpha_{12})$ is in general non-zero if $P_{qi} (q=1,2)$ have $\alpha$-dependences 
even if $c_{ij}^{(12)}(\alpha_1,\alpha_2) =0$ so that $\hat\rho^{(12)}(\alpha_1,\alpha_2)=\hat\rho^{(1)}(\alpha_1)\otimes\hat\rho^{(2)}(\alpha_2)$. 
To distinguish them from each other, we propose to call $c_{ij}^{(12)}(\alpha_1,\alpha_2)$ the genuine spin correlation
but the corresponding $\bar c_{ij}^{(12)}(\alpha_{12})$ the effective spin correlation 
and $\bar c_{ij}^{(12;0)}(\alpha_{12})$ the induced spin correlation. 
To be consistent, we will also call $\bar P_{qi}(\alpha_{12})$ the effective and $P_{qi}(\alpha_{q})$ the genuine polarization.  

Also we suggest to distinguish the induced spin correlations into
local and long range correlations depending on whether they are short
or long ranged in the $\alpha$-space. An example that leads to such
induced spin correlations was given in Refs.~\citep{Yang:2017sdk,Sheng:2019kmk,Sheng:2020ghv,Sheng:2022ffb,Sheng:2022wsy}.
The spin correlation between $s$ and $\bar{s}$ was shown to be strong
and local in phase space due to strong interaction exchanged $\phi$-meson field.

Similarly, for a three-particle system (123), the $\alpha$-dependent spin density matrix reads 
\begin{align}
\hat\rho^{(123)}(\alpha_1,\alpha_2,\alpha_3)=&\hat\rho^{(1)}(\alpha_1)\otimes \hat\rho^{(2)}(\alpha_2)\otimes \hat\rho^{(3)}(\alpha_3) \nonumber\\
+&\frac{1}{2^2} \Bigl[ c_{ij}^{(12)}(\alpha_1,\alpha_2) \hat\sigma_{1i}\otimes\hat\sigma_{2j}\otimes \hat\rho^{(3)}(\alpha_3)
+c_{jk}^{(23)}(\alpha_2,\alpha_3) \hat\rho^{(1)}(\alpha_1) \otimes\hat\sigma_{2j}\otimes\hat\sigma_{3k} \nonumber\\
&+c_{ik}^{(13)}(\alpha_1,\alpha_3) \hat\sigma_{1i}\otimes\hat\rho^{(2)}(\alpha_2)\otimes\hat\sigma_{3k} \Bigr]
+\frac{1}{2^3}c_{ijk}^{(123)}(\alpha_1,\alpha_2,\alpha_3) \hat\sigma_{1i}\otimes \hat\sigma_{2j}\otimes\hat\sigma_{3k}.  \label{eq:rho123b}
\end{align} 
If the system (123) is in the state $\left|\alpha_{123}\right\rangle $,
the effective spin density matrix are given by 
\begin{align}
\hat{\bar\rho}^{(123)}(\alpha_{123})=&\hat{\bar\rho}^{(1)}(\alpha_{123})\otimes \hat{\bar\rho}^{(2)}(\alpha_{123})\hat{\bar\rho}^{(3)}(\alpha_{123})\nonumber\\
&+\frac{1}{2^2} \Bigl[\bar c_{ij}^{(12)}(\alpha_{123}) \hat\sigma_{1i}\otimes\hat\sigma_{2j}\otimes \hat{\bar\rho}^{(3)}(\alpha_{123})
+\bar c_{jk}^{(23)}(\alpha_{123}) \hat{\bar\rho}^{(1)}(\alpha_{123}) \otimes\hat\sigma_{2j}\otimes\hat\sigma_{3k}  \nonumber\\
&+\bar c_{ik}^{(13)}(\alpha_{123}) \hat\sigma_{1i}\otimes\hat{\bar\rho}^{(2)}(\alpha_{123})\otimes\hat\sigma_{3k} \Bigr]
+\frac{1}{2^3}\bar c_{ijk}^{(123)}(\alpha_{123}) \hat\sigma_{1i}\otimes\hat\sigma_{2j}\otimes\hat\sigma_{3k} , \label{eq:rho123b2}
\end{align} 
where the effective polarizations such as $\bar{P}_{1i}(\alpha_{123})$ 
and effective two-particle correlations such as $\bar{c}_{ij}^{(12)}(\alpha_{123})$
have similar expressions as those in the two-particle system given
by Eqs.~(\ref{eq:bP1i}) and (\ref{eq:bc12}), and the effective three-particle
correlation $\bar{c}_{ijk}^{(123)}(\alpha_{123})$ is given by 
\begin{align}
\bar c_{ijk}^{(123)} (\alpha_{123}) = &\langle  c_{ijk}^{(123)}(\alpha_1,\alpha_2,\alpha_3) +P_{1i} (\alpha_{1})P_{2j} (\alpha_{2})P_{3k} (\alpha_{3}) \nonumber\\
&+ c_{ij}^{(12)}(\alpha_{1},\alpha_2) P_{3k}(\alpha_3) + c_{ik}^{(13)}(\alpha_{1},\alpha_3) P_{2j}(\alpha_2) + c_{jk}^{(23)}(\alpha_{2},\alpha_3) P_{1i}(\alpha_1) \rangle   \nonumber\\
& - \bar c_{ij}^{(12)}(\alpha_{123}) \bar P_{3k}(\alpha_{123}) - \bar c_{ik}^{(13)}(\alpha_{123}) \bar P_{2j}(\alpha_{123}) - \bar c_{jk}^{(23)}(\alpha_{123}) \bar P_{1i}(\alpha_{123}) \nonumber\\
&- \bar P_{1i} (\alpha_{123}) \bar P_{2j} (\alpha_{123})\bar P_{3k} (\alpha_{123}). \label{eq:bc123}
\end{align}

If $c_{ij}^{(12)}(\alpha_{1},\alpha_{2},\alpha_{3})=0$ and $c_{ijk}^{(123)}(\alpha_{1},\alpha_{2},\alpha_{3})=0$,
the spin density matrix of the system in Eq.~(\ref{eq:rho123b}) is the direct product of single-particle spin density matrices. 
In this case we have a similar result for the induced two-particle spin correlations
to Eq.~(\ref{eq:bc120}), and the induced three-particle spin correlation becomes
\begin{align}
\bar c_{ijk}^{(123;0)} (\alpha_{123})=& \langle P_{1i} (\alpha_{1})P_{2j} (\alpha_{2}) P_{3k} (\alpha_{3})\rangle
+2 \langle P_{1i} (\alpha_{1})\rangle \langle P_{2j} (\alpha_2)\rangle  \langle P_{3k} (\alpha_{3})\rangle \nonumber\\
& -\langle P_{1i}(\alpha_1)P_{2j}(\alpha_2)\rangle \langle P_{3k}(\alpha_{3}) \rangle 
-\langle  P_{1i}(\alpha_1)P_{3k}(\alpha_3) \rangle \langle P_{2j}(\alpha_{2}) \rangle 
-\langle  P_{2j}(\alpha_2)P_{3k}(\alpha_3) \rangle \langle P_{1i}(\alpha_{1}) \rangle, \label{eq:bc1230} 
\end{align}
If $c_{ijk}^{(123)}(\alpha_{1},\alpha_{2},\alpha_{3})=0$ but there
are two-particle spin correlations, the induced three-particle spin
correlation has the form 
\begin{align}
\bar c_{ijk}^{(123;1)} (\alpha_{123}) = &\langle P_{1i} (\alpha_{1}) P_{2j} (\alpha_{2}) P_{3k} (\alpha_{3}) 
+ c_{ij}^{(12)}(\alpha_{1},\alpha_2) P_{3k}(\alpha_3) + c_{ik}^{(13)}(\alpha_{1},\alpha_3) P_{2j}(\alpha_2) + c_{jk}^{(23)}(\alpha_{2},\alpha_3) P_{1i}(\alpha_1) \rangle   \nonumber\\
& - \bar c_{ij}^{(12)}(\alpha_{123}) \bar P_{3k}(\alpha_{123}) - \bar c_{ik}^{(13)}(\alpha_{123}) \bar P_{2j}(\alpha_{123}) - \bar c_{jk}^{(23)}(\alpha_{123}) \bar P_{1i}(\alpha_{123}) \nonumber\\ 
& - \bar P_{1i} (\alpha_{123}) \bar P_{2j} (\alpha_{123})\bar P_{3k} (\alpha_{123}). \label{eq:bc1231}
\end{align}
We see that if the system only has two-particle spin correlations
but has $\alpha$-dependence, three-particle spin correlations are
not vanishing due to averages over $\alpha$ in a given region and/or
given $\alpha$-dependent weight. 
We call $\bar{c}_{ijk}^{(123;1)}(\alpha_{123})$ in Eq.~(\ref{eq:bc1231}) the first order induced spin correlation
and $\bar{c}_{ijk}^{(123;0)}(\alpha_{123})$ in Eq.~(\ref{eq:bc1230}) the zeroth order induced spin correlation. 

\section{Spin density matrix for vector mesons}\label{sec:Vpol}

We take a simple case that quarks and anti-quarks in the system combine
with each other to form hadrons. We use this as an illustrating example
to show the relationship between the quark-quark spin correlations
and the polarization of hadrons as well as other measurable quantities. 

In this section, we consider the combination process $q_{1}+\bar{q}_{2}\to V$
and present the results for the spin density matrix of the vector
meson $V$. We use $\hat{\mathcal M}$ to denote the transition matrix
for a $q_{1}\bar{q}_{2}$ to combine together to form $V$ in the
combination process so that the spin density matrix of $V$ is given by 
\begin{align}
\hat\rho^{V}=\hat{\cal M}\hat\rho^{(q_1\bar q_2)}\hat{\cal M}^\dag. \label{eq:rhoV}
\end{align}
Using this, we will calculate elements of $\hat{\rho}^{V}$ in various cases in this section. 

\subsection{With only spin degree of freedom \label{sec:rhoV}}

If we only consider the spin degree of freedom, the matrix element of $\hat{\rho}^{V}$ is given by 
\begin{align}
\rho^{V}_{mm'}=\langle jm|\hat{\cal M}\hat\rho^{(q_1\bar q_2)} \hat{{\cal M}}^\dag | jm'\rangle
=\sum_{m_n,m'_n} \langle jm|\hat{\cal M}|m_n\rangle\langle m_n|\hat\rho^{(q_1\bar q_2)}|m'_n\rangle\langle m'_n|\hat{\cal M}^\dag| jm'\rangle , \label{eq:rhoV1}
\end{align}
where $j=1$ is the spin of $V$. 
Hereafter, we will use shorthand notations $|m_{n}\rangle \equiv |j_{1}m_{1},j_{2}m_{2}\rangle$
and $|m_{n}^{\prime}\rangle \equiv |j_{1}m_{1}^{\prime},j_{2}m_{2}^{\prime}\rangle$
for spin states of quark-antiquark system in case of no ambiguity. 

The transition matrix element $\langle jm|\hat{\cal M}|m_{n}\rangle$ can be further written as 
\begin{align}
\langle jm|\hat{\cal M}|m_n\rangle=\sum_{j'm'} \langle jm|\hat{\cal M}|j'm'\rangle\langle j'm'|m_n\rangle, \label{eq:tmV1}
\end{align}
where $\left\langle m_n|jm\right\rangle $ is the well-known Clebsch-Gordan coefficient. 
The space rotation invariance demands that $j=j^{\prime}$ and $m=m^{\prime}$ and 
that $\left\langle jm|\hat{\cal M}|jm\right\rangle$ be independent of $m$. 
We therefore obtain that 
\begin{align}
\rho^{V}_{mm'}=N_V\sum_{m_n;m_n'} \langle jm|m_n\rangle \langle m_n|\hat\rho^{(q_1\bar q_2)}|m'_n\rangle \langle m'_n|jm'\rangle, \label{eq:rhoV2}
\end{align}
where $N_{V}$ is a constant that can be absorbed into the normalization constant.

We insert $\hat{\rho}^{(q_{1}\bar{q}_{2})}$ by Eq.~(\ref{eq:rho12})
into Eq. (\ref{eq:rhoV2}) and obtain the element of the vector meson's
spin density matrix 
\begin{align}
\rho_{00}^{V}=&\frac{1}{C_V} \Bigl\{1+c_{xx}^{(q_1\bar q_2)}+c_{yy}^{(q_1\bar q_2)}-c_{zz}^{(q_1\bar q_2)}+P_{q_1x} P_{\bar{q}_2x}+P_{q_1y} P_{\bar{q}_2y}-P_{q_1z} P_{\bar{q}_2z}\Bigr\},  \label{eq:rho00} \\
\rho^V_{1-1}=&\frac{1}{C_V} \Bigl\{c_{xx}^{(q_1\bar q_2)}-c_{yy}^{(q_1\bar q_2)}+P_{q_1x}P_{\bar{q}_2x}-P_{q_1y}P_{\bar{q}_2y} -i[c_{xy}^{(q_1\bar q_2)}+c_{yx}^{(q_1\bar q_2)} +P_{q_1x}P_{\bar{q}_2y} + P_{q_1y} P_{\bar{q}_2x}] \Bigr\},  \label{eq:rho1m1} \\ 
\rho_{10}^{V}=&\frac{1}{\sqrt{2} C_V}  \Bigl\{c_{xz}^{(q_1\bar q_2)}+c_{zx}^{(q_1\bar q_2)}+P_{q_1x}(1+P_{\bar{q}_2z}) +( 1+P_{q_1z}) P_{\bar{q}_2x} \nonumber\\
& -i[ c_{yz}^{(q_1\bar q_2)}+c_{zy}^{(q_1\bar q_2)}+ P_{q_1y} ( 1+P_{\bar{q}_2z}) +( 1+P_{q_1z}) P_{\bar{q}_2y}]\Bigr\}, \label{eq:rho10} \\
\rho_{0-1}^{V}=&\frac{1}{\sqrt{2} C_V} \Bigl\{-c_{xz}^{(q_1\bar q_2)}-c_{zx}^{(q_1\bar q_2)}+P_{q_1x} (1-P_{\bar{q}_2z}) + (1-P_{q_2z})P_{\bar{q}_2x} \nonumber\\
&+i[c_{yz}^{(q_1\bar q_2)}+c_{zy}^{(q_1\bar q_2)} - P_{q_1y} (1-P_{\bar{q}_2z} ) - P_{\bar{q}_2y} (1-P_{q_1z})] \Bigr\}, \label{eq:rho0m1} 
\end{align}
where $C_V= {\rm Tr}\hat\rho^V=3+c_{ii}^{(q_1\bar q_2)}+{P}_{q_1i}{P}_{\bar{q}_2i}$ is the normalization constant.

From Eqs.~(\ref{eq:rho00}-\ref{eq:rho0m1}),  we see clearly that we have contributions from quark-anti-quark spin correlations 
in all elements of the spin density matrix of the vector meson.

\subsection{With other degrees of freedom\label{sec:rhoValpha}}

If there are other degrees of freedom, we have 
\begin{align}
\rho^{V}_{mm'}(\alpha_V)=&\langle jm,\alpha_V|\hat{\cal M}\hat\rho^{(q_1\bar q_2)}\hat{{\cal M}}^\dag | jm',\alpha_V\rangle \nonumber\\
=&\sum_{m_n;m'_n} \sum_{\alpha_n} \langle jm,\alpha_V|\hat{\cal M}|m_n,\alpha_n\rangle\langle m_n|\hat\rho^{(q_1\bar q_2)}(\alpha_n)|m'_n\rangle\langle m'_n,\alpha_n|\hat{{\cal M}}^\dag| jm',\alpha_V\rangle , \label{eq:rhoV1a}
\end{align}
where similar to $m_n$, we use $\alpha_n$ to denote $(\alpha_1,\alpha_2)$; 
and we consider only the case discussed in Sec.~\ref{sec:qcalpha},
i.e. we consider only the $\alpha$-dependence of $\hat{\rho}^{(q_{1}\bar{q}_{2})}$
but neglect its off-diagonal elements with respect to $\alpha$. 

The transition matrix element $\langle jm,\alpha_V|\hat{\cal M}|m_n,\alpha_n\rangle$ is further simplified as, 
\begin{align}
\langle jm,\alpha_V|\hat{\cal M}|m_n,\alpha_n\rangle 
=\sum_{\alpha'_V,j'm'} \langle jm,\alpha_V|\hat{\cal M}|j'm',\alpha'_V\rangle\langle j'm',\alpha'_V|m_n,\alpha_n\rangle. \label{eq:rhoV1a2}
\end{align}
If we consider only the case where all $j,m$ and $\alpha$ are conserved, we obtain
\begin{align}
\langle jm,\alpha_V|\hat{\cal M}|m_n,\alpha_n\rangle=\langle jm,\alpha_V|\hat{\cal M}|jm,\alpha_V\rangle\langle jm,\alpha_V|m_n,\alpha_n\rangle. 
\end{align}
where the rotation invariance of $\hat{\mathcal{M}}$ leads to that $\langle jm,\alpha_{V}|\hat{\mathcal{M}}|jm,\alpha_{V}\rangle$
is independent of $m$. In this case, the matrix element $\rho_{mm^{\prime}}^{V}(\alpha_{V})$
is obtained as 
\begin{align}
\rho^{V}_{mm'}(\alpha_V) 
=&N(\alpha_V) \sum_{m_n;m'_n}  \langle jm|m_n\rangle\langle m'_n|jm'\rangle \langle m_n|\hat{\bar\rho}^{(q_1\bar q_2)}(\alpha_V)|m'_n\rangle, \label{eq:rhoV1b} 
\end{align}
where $N_{V}(\alpha_{V})$ is a constant for given $\alpha_V$  and $\hat{\bar{\rho}}^{(q_{1}\bar{q}_{2})}(\alpha_{V})$ is given by 
\begin{align}
&\hat{\bar\rho}^{(q_1\bar q_2)}(\alpha_V)
= \sum_{\alpha_n} \hat{\rho}^{(q_1\bar q_2)}(\alpha_n)\psi^*(\alpha_n;\alpha_V;m,m_n)\psi(\alpha_n;\alpha_V;m',m'_n),
\end{align}
that in general depends on $m$, $m^{\prime}$, $m_n$ and $m_n^{\prime}$ through the function 
\begin{equation}
\psi(\alpha_n;\alpha_V;m,m_n)=\frac{ \langle m_n,\alpha_n|jm,\alpha_V\rangle}{ \langle m_n|jm\rangle}. 
\end{equation}
Note that $\psi$ does not depend on $m$ and $m_n$ if the wave function is factorized, i.e, 
\begin{align}
&\langle m_n,\alpha_n|jm,\alpha_V\rangle=\langle\alpha_n|\alpha_V\rangle\langle m_n|jm\rangle. \label{eq:fact1} 
\end{align}
In this case, we have $\psi(\alpha_n;\alpha_V;m',m'_n)= \langle \alpha_n|\alpha_V\rangle$ that depends on $(\alpha_n,\alpha_V)$ only  and 
\begin{align}
&\hat{\bar\rho}^{(q_1\bar q_2)}(\alpha_V)=\langle \hat{\rho}^{(q_1\bar{q}_2)}(\alpha_n) \rangle _V
\equiv \sum_{\alpha_n} \hat{\rho}^{(q_1\bar q_2)}(\alpha_n) |\langle \alpha_n|\alpha_V\rangle|^2. \label{eq:fact3}
\end{align}
We have a similar formula to Eq.~(\ref{eq:rho12b}) for $\hat{\bar{\rho}}^{(q_{1}\bar{q}_{2})}(\alpha_{V})$, i.e., 
\begin{equation}
\hat{\bar{\rho}}^{(q_{1}\bar{q}_{2})}(\alpha_{V})=\hat{\bar{\rho}}^{(q_1)}(\alpha_{V})\otimes\hat{\bar{\rho}}^{(\bar{q}_2)}(\alpha_{V})
+\frac{1}{2^{2}}\bar{c}_{ij}^{(q_1\bar{q}_2)}(\alpha_{V})\hat{\sigma}_{1i}\otimes\hat{\sigma}_{2j},\label{eq:fact4}
\end{equation}
where the effective spin density of $q_1$ and the effective spin correlation are given by 
\begin{align}
\hat{\bar{\rho}}^{(q_1)}(\alpha_{V}) = & \left\langle \hat{\rho}^{(q_1)}(\alpha_{1})\right\rangle _{V}
=\sum_{\alpha_n} |\langle\alpha_n|\alpha_{V}\rangle|^2 \hat{\rho}^{(q_1)}(\alpha_{1})
=  \frac{1}{2}[1+\bar{P}_{q_1i}(\alpha_{V})\hat{\sigma}_{1i}],\nonumber \\
\bar{P}_{q_1i}(\alpha_{V})= & \left\langle P_{q_1i}(\alpha_{1})\right\rangle _{V}=\sum_{\alpha_n} 
|\langle\alpha_n|\alpha_{V}\rangle|^2 P_{q_1i}(\alpha_{1}),\nonumber \\
\bar{c}_{ij}^{(q_1\bar{q}_2)}(\alpha_{V})= & \left\langle c_{ij}^{(q_1\bar{q}_2)}(\alpha_n)+P_{qi}(\alpha_{1})P_{\bar{q}j}(\alpha_{2})\right\rangle _{V}-\left\langle P_{q_1i}(\alpha_{1})\right\rangle _{V}\left\langle P_{\bar{q}_2j}(\alpha_{2})\right\rangle _{V}, \label{eq:rho-o-c-v}
\end{align}
and similar for $\hat{\bar{\rho}}^{(\bar q_2)}(\alpha_{V})$ and $\bar{P}_{{\bar q}_2i}(\alpha_{V})$.  
We see that the above results are similar to Eqs. (\ref{eq:bP1i}) and (\ref{eq:bc12}).

Eq.~(\ref{eq:fact3}) just corresponds to the case discussed in Sec.~\ref{sec:qcalpha}. 
We note that the factorization in Eq.~(\ref{eq:fact1}) is true in non-relativistic quark models. In relativistic
quantum systems, spin is not an independent degree of freedom so the
wave function is not factorizable as in Eq.~(\ref{eq:fact1}). 
In the following of this paper, we limit ourselves to the factorizable case and leave the general case for future studies.

Using Eqs.~(\ref{eq:rhoV1b}) and (\ref{eq:fact4}), we can obtain
the results for $\rho_{mm^{\prime}}^{V}(\alpha_{V})$ similar to $\rho_{mm^{\prime}}^{V}$ in Eqs.~(\ref{eq:rho00}-\ref{eq:rho0m1}). 
The results can be obtained from Eqs.~(\ref{eq:rho00}-\ref{eq:rho0m1})
by the replacement of spin polarization and correlation quantities, $P_{q_1i}\rightarrow\bar{P}_{q_1i}(\alpha_{V})$,
$P_{\bar{q}_2j}\rightarrow\bar{P}_{\bar{q}_2j}(\alpha_{V})$ and
$c_{ij}^{(q_1\bar{q}_2)}\rightarrow\bar{c}_{ij}^{(q_1\bar{q}_2)}(\alpha_{V})$,
where $\bar{P}_{q_1i}(\alpha_{V})$ and $\bar{c}_{ij}^{(q_1\bar{q}_2)}(\alpha_{V})$ are defined in Eq.~(\ref{eq:rho-o-c-v}).

By applying Eq.~(\ref{eq:bc12}), the above results can be put in
similar forms as those given by Eqs.~(\ref{eq:rho00}-\ref{eq:rho0m1}) but with the average
taken over each numerator and each denominator separately weighted by $|\langle\alpha_n|\alpha_{V}\rangle|^2$.
Here we show the result of $\rho_{00}^{V}$ corresponding to Eq.~(\ref{eq:rho00}),  
\begin{equation}
\rho_{00}^{V}(\alpha_V)=\frac{1}{\langle C_{V}\rangle_{V}} 
\left[1+\langle P_{q_1x}P_{\bar{q}_2x}+P_{q_1y}P_{\bar{q}_2y}-P_{q_1z}P_{\bar{q}_2z}
+c_{xx}^{(q_1\bar{q}_2)}+c_{yy}^{(q_1\bar{q}_2)}-c_{zz}^{(q_1\bar{q}_2)}\rangle_{V}\right].\label{eq:rhoV00a2}
\end{equation}

In practice, in particular in experiments, we often study $\hat{{\rho}}^V(\alpha_{V})$ 
averaged over $\alpha_{V}$ in a given kinematic region. 
In this case, we obtain e.g for $\rho^V_{00}$,   
\begin{align}
\langle \rho _{00}^{V}\rangle =&\frac{1}{\langle\langle C_{V}\rangle_{V}\rangle_S} 
\left[1+\langle \bar P_{q_1x}\bar P_{\bar{q}_2x}+\bar P_{q_1y}\bar P_{\bar{q}_2y}-\bar P_{q_1z}\bar P_{\bar{q}_2z}
+\bar c_{xx}^{(q_1\bar{q}_2)}+\bar c_{yy}^{(q_1\bar{q}_2)}-\bar c_{zz}^{(q_1\bar{q}_2)}\rangle_S\right] \nonumber\\
=&\frac{1}{\langle\langle C_{V}\rangle_{V}\rangle_S} 
\left[1+\langle\langle P_{q_1x}P_{\bar{q}_2x}+P_{q_1y}P_{\bar{q}_2y}-P_{q_1z}P_{\bar{q}_2z}
+c_{xx}^{(q_1\bar{q}_2)}+c_{yy}^{(q_1\bar{q}_2)}-c_{zz}^{(q_1\bar{q}_2)}\rangle_{V}\rangle_S\right], 
\label{eq:arhoV00a2} 
\end{align}
where $S$ denotes the kinematic region of $\alpha_V$ or the sub-system over that we average. 

We emphasized~\cite{talks} in particular that the average now is two folded. 
For example, for $c_{ij}^{(q_1\bar{q}_2)}$, it is 
\begin{align}
\langle c_{ij}^{(q_1\bar{q}_2)}\rangle =\sum_{\alpha_V} f_V(\alpha_V) \langle c_{ij}^{(q_1\bar{q}_2)}(\alpha_n) \rangle_V
=\sum_{\alpha_V} f_V(\alpha_V) \sum_{\alpha_n} |\langle\alpha_n |\alpha_V\rangle|^2 
c_{ij}^{(q_1\bar{q}_2)}(\alpha_n) , \label{eq:2faverage}
\end{align}
where $f_V(\alpha_V)$ is the $\alpha_V$ distribution of $V$. 
We see that for the genuine quark spin correlation $c_{ij}^{(q_1\bar q_2)}(\alpha_{q_1}, \alpha_{\bar q_2})$ 
and polarizations $P_{q_1}(\alpha_{q_1})$ and $P_{\bar q_2}(\alpha_{\bar q_2})$ of $q_1$ and $\bar q_2$, 
we first average over $(\alpha_{q_1}, \alpha_{\bar q_2})$ inside the vector meson $V$. 
In this step, we obtain $\langle c_{ij}^{(q_1\bar q_2)}(\alpha_{q_1}, \alpha_{\bar q_2})\rangle_V$
and the induced correlation 
$c_{ij}^{(q_1\bar q_2,0)}(\alpha_V)=\langle P_{q_1}(\alpha_{q_1})P_{\bar q_2}(\alpha_{\bar q_2})\rangle_V
-\langle P_{q_1}(\alpha_{q_1})\rangle_V\langle P_{\bar q_2}(\alpha_{\bar q_2})\rangle_V$.  
It is clear that in this step only local quark-anti-quark spin correlations contribute. 
In the second step, we average over $V$ with different $\alpha_V$. 
For both the genuine and induced correlations,  we average the results obtained in the first step at different $\alpha_V$ 
weighted by the $\alpha_V$ distribution $f_V(\alpha_V)$ of $V$. 
We never consider a $q_1$ and $\bar q_2$ separated by a large distance in the $\alpha$-space. 
Hence we do not have any contribution from long range correlation. 
This implies that by studying vector meson spin alignment and off-diagonal elements of the spin density matrix, 
we study only local quark-anti-quark spin correlations.

\section{Global hyperon polarization \label{sec:Hpol}}

For hyperon formation from three quarks, $q_{1}+q_{2}+q_{3}\to H$,
similar to the vector meson discussed in Sec.~\ref{sec:rhoV}, the
spin density matrix for the hyperon is given by 
\begin{align}
\hat\rho^H=\hat{\cal M}\hat\rho^{(q_1q_2q_3)}\hat{\cal M}^\dag. \label{eq:rhoH}
\end{align}
The calculation of the hyperon's polarization is also similar to the
vector meson which we will present in this section. 

\subsection{With only spin degrees of freedom}

In this case, the matrix element of $\rho_{mm^{\prime}}^{H}$ is written as 
\begin{align}
\rho^H_{m m^{\prime}}=&\langle{jm} | \hat{\mathcal{M}} \hat{\rho}^{(q_1q_2q_3)}{\hat{\mathcal{M}}}^{\dagger}| {j m^{\prime}}\rangle 
=\sum_{m_n;m'_n}\langle{jm}|
\hat{\mathcal{M}} |{m_n}\rangle \langle{m_n}|\hat{\rho}^{(q_1q_2q_3)}|{m_n^\prime}\rangle \langle{m_n^{\prime}}| \hat{\mathcal{M}}^{\dagger} |{jm^\prime}\rangle,
\end{align}
where $j={1}/{2}$ is the spin of the hyperon $H$. 
For the quark spin state, we still omit $j_i$ in $|j_im_n\rangle$ since they are all $1/2$ and use the shorthand notation $|m_n\rangle$ 
to stand for $|m_1,m_2,m_3\rangle$. 
Similar to Eq.~(\ref{eq:tmV1}), the transition matrix element $\langle {jm}|\hat{\mathcal{M}}|{m_n}\rangle$ is given by
\begin{equation}
\langle {jm}|\hat{\mathcal{M}}|{m_n}\rangle=\sum_{j^{\prime} m^{\prime}}
\langle{jm}|\hat{\mathcal{M}}|{j^{\prime}m^{\prime}}\rangle \langle{j^{\prime}m^{\prime}}| {m_n} \rangle.   \label{eq:MHmm}
\end{equation}
We use again the space rotation invariance 
that demands that $j=j'$ and $m=m'$ and that $\langle jm|\hat{\cal M}|jm\rangle$ is independent of $m$. 
We therefore obtain that 
\begin{equation}
\rho^H_{m m^\prime}=N_H \sum_{m_n;m'_n}
\langle jm|m_n\rangle \langle m_n|\hat{\rho}^{(q_1q_2q_3)}|m'_n\rangle \langle m'_n|jm'\rangle. \label{eq:rhoHmm}
\end{equation}

Note in particular that the spin density matrix in Eq.~(\ref{eq:rhoHmm}) for the
hyperon has the same form as that in Eq.~(\ref{eq:rhoV2}) for the vector meson. 

By inserting Eq.~(\ref{eq:rho123}) into Eq.~(\ref{eq:rhoHmm}),
we obtain the spin density matrix $\hat{\rho}^{H}$ and the hyperon
polarization $P_{H}=\Delta\rho_{H}/C_{H}$ where $\Delta \rho_{H}=\rho_{1/2,1/2}^{H}-\rho_{-1/2,-1/2}^{H}$
and $C_{H}={\rm Tr}\hat\rho^H$ is the normalization constant. 
The result for $\varLambda$ is given by
\begin{align}
& P_{\varLambda}=P_{sz}-\frac{\delta \rho_\varLambda}{C_\varLambda}, \label{eq:PLambda} \\
& \delta \rho_{\varLambda}=c_{iz}^{(us)}P_{di}+c_{iz}^{(ds)}P_{ui}+c_{iiz}^{(uds)}, \label{eq:dALambda} \\
& C_{\varLambda}=1-c_{ii}^{(ud)}-P_{ui}P_{di}. \label{eq:CLambda}
\end{align}
From Eqs.~(\ref{eq:PLambda}-\ref{eq:CLambda}), we see in particular that the $\varLambda$ polarization  
is not simply equal to that of the $s$-quark when quark spin correlations are considered. 
This is because to produce a spin-up or spin-down $\varLambda$, 
we need not only a spin-up or spin-down $s$-quark but also a spin zero $ud$-di-quark. 
 If the spin of $s$ and those of $u$ and $d$ are correlated, the probability to have a spin zero $ud$-di-quark 
 in the case that $s$ is spin-up can be different from that in the case that $s$ is spin-down. 
 This leads to influences on the final polarization of $\varLambda$. 

For other $J^{P}=(1/2)^{+}$ hyperons, we obtain 
\begin{align}
P_{H_{112}}= & \frac{1}{3}(4P_{q_1z}-P_{q_2z})+\frac{\delta \rho_{H_{112}}}{C_{H_{112}}},\label{eq:PH112} \\
P_{\Sigma^{0}}= & \frac{1}{3}(2P_{uz}+2P_{dz}-P_{sz})+\frac{\delta \rho_{\Sigma^{0}}}{C_{\Sigma^{0}}}, \label{eq:Psigma0}
\end{align}
where $\delta \rho_{H_{112}}$, $C_{H_{112}}$, $\delta \rho_{\Sigma^{0}}$ and $C_{\Sigma^{0}}$ are given by 
\begin{align}
\delta \rho_{H_{112}}= & -\frac{4}{3}\left(P_{q_1i}P_{q_1i}-P_{q_1i}P_{q_2i}+c_{ii}^{(q_1q_1)}-c_{ii}^{(q_1q_2)}\right)\left(P_{q_1z}-P_{q_2z}\right)\nonumber \\
 & -4c_{iz}^{(q_1q_1)}P_{q_2i}+2\left(c_{iz}^{(q_1q_2)}-2c_{zi}^{(q_1q_2)}\right)P_{q_1i}+c_{iiz}^{(q_1q_1q_2)}-4c_{zii}^{(q_1q_1q_2)},\label{eq:dAHq1q1q2}\\
C_{H_{112}}= & 3+P_{q_1i}P_{q_1i}-4P_{q_1i}P_{q_2i}+c_{ii}^{(q_1q_1)}-4c_{ii}^{(q_1q_2)},\label{eq:CHq_1q_1q_2}\\
\delta \rho_{\Sigma^{0}}= & -\frac{2}{3}\left(P_{ui}P_{di}+c_{ii}^{(ud)}\right)\left(P_{uz}+P_{dz}-2P_{sz}\right)
+\frac{2}{3}\left(P_{ui}P_{si}+c_{ii}^{(us)}\right)\left(2P_{uz}-P_{dz}-P_{sz}\right)\nonumber \\
 & +\frac{2}{3}\left(P_{di}P_{si}+c_{ii}^{(ds)}\right)\left(2P_{dz}-P_{uz}-P_{sz}\right)+\left(c_{zi}^{(us)}-2c_{iz}^{(us)}\right)P_{di}\nonumber \\
 & +\left(c_{zi}^{(ds)}-2c_{iz}^{(ds)}\right)P_{ui}-2\left(c_{iz}^{(ud)}+c_{zi}^{(ud)}\right)P_{si}+c_{iiz}^{(uds)}-2c_{izi}^{(uds)}-2c_{zii}^{(uds)},\label{eq:dAsigma0}\\
C_{\Sigma^{0}}= & 3+P_{ui}P_{di}-2P_{di}P_{si}-2P_{ui}P_{si}+c_{ii}^{(ud)}-2c_{ii}^{(us)}-2c_{ii}^{(ds)},\label{eq:Csigma0}
\end{align}
where $H_{112}$ denotes $J^{P}=(1/2)^{+}$ hyperon with quark flavor
$q_1q_1q_2$ and we have used $c_{xz}^{(q_1q_1)}=c_{zx}^{(q_1q_1)}$ and $c_{ijk}^{(q_1q_1q_2)}=c_{jik}^{(q_1q_1q_2)}$.
We see again that there are contributions from spin correlations in hyperons' polarizations.

\subsection{With other degrees of freedom}

If there are other degrees of freedom besides spin, the matrix elements
of $\rho_{mm^{\prime}}^{H}(\alpha_{H})$ is given by 
\begin{align}
\rho^H_{mm^{\prime}}(\alpha_H)
=&\langle{jm,\alpha_H}|\hat{\mathcal{M}}\hat{\rho}^{(q_1q_2q_3)}{(\alpha_n)}{\hat{\mathcal{M}}}^{\dagger}|{j m^\prime,\alpha_H}\rangle \nonumber\\
=&\sum_{m_n,m'_n}\sum_{\alpha_n}
\langle{jm,\alpha_H}|\hat{\mathcal{M}}|{m_n,\alpha_n}\rangle \langle{m_n^\prime,\alpha_n}|\hat{\mathcal{M}}^{\dagger}|{jm^{\prime},\alpha_H}\rangle 
 \langle{m_n}|\hat{\rho}^{(q_1q_2q_3)}(\alpha_n)|{m_n^\prime}\rangle.    \label{eq:rhoHmma}
 \end{align}
The transition matrix element $\langle jm,\alpha_{H}|\mathcal{M}|m_n,\alpha_n\rangle$ can be further written as 
\begin{align}
\langle jm,\alpha_H|\hat{\cal M}|m_n,\alpha_n\rangle
=\sum_{\alpha^{\prime}_H,j^{\prime}m^{\prime}} \langle jm,\alpha_H|\hat{\cal M}|j^{\prime}m^{\prime},\alpha^{\prime}_H\rangle\langle j^{\prime}m^{\prime},\alpha^{\prime}_H|m_n,\alpha_n\rangle.  \label{eq:MHmmia}
\end{align}
In the case that $j,m$ and $\alpha$ are conserved, we obtain 
\begin{align}
\langle jm,\alpha_H|\hat{\cal M}|m_n,\alpha_n\rangle=\langle jm,\alpha_H|\hat{\cal M}|jm,\alpha_H\rangle\langle jm,\alpha_H|m_n,\alpha_n\rangle,  \label{eq:MHmmia2}
\end{align}
so that the matrix element $\rho^{H}_{mm^{\prime}}(\alpha_H)$ is given by
\begin{align}
\rho^{H}_{mm^{\prime}}(\alpha_H)
=&N_H(\alpha_H) \sum_{m_n,m_n^{\prime}}  \langle jm|m_n\rangle\langle m'_n|jm^{\prime}\rangle \langle m_n|\hat{\bar\rho}^{(q_1 q_2 q_3)}(\alpha_H)|m'_n\rangle, 
\label{eq:rhoHmma2} 
\end{align}
where $\hat{\bar\rho}^{(q_1 q_2 q_3)}(\alpha_H)$ in general depends on $m,m_n,m^{\prime},m'_n$ and is given by
\begin{align}
&\hat{\bar\rho}^{(q_1 q_2 q_3)}(\alpha_H)= \sum_{\alpha_n} \hat{\rho}^{(q_1 q_2 q_3)}(\alpha_n)\psi^*(\alpha_n,\alpha_H;m,m_n)
\psi(\alpha_n;\alpha_H;m^{\prime},m'_n), \label{eq:brho3qa}. 
\end{align}
where $\psi(m_n,m;\alpha_n,\alpha_{H})$ is defined as
\begin{equation}
\psi(m_n,m;\alpha_n,\alpha_{H})=\frac{\langle m_n,m,\alpha_n|jm,\alpha_{H}\rangle}{\langle m_n|jm\rangle}.\label{eq:psia123}
\end{equation}
If the wave function is factorizable, $\langle m_n,\alpha_n|jm,\alpha_{H}\rangle=\langle\alpha_n|\alpha_{H}\rangle\langle m_n|jm\rangle$,
we have $\psi(m_n,m;\alpha_n,\alpha_{H})=\langle\alpha_n|\alpha_{H}\rangle$.
So Eq.~(\ref{eq:brho3qa}) is simplified as 
\begin{align}
&\hat{\bar\rho}^{(q_1 q_2 q_3)}(\alpha_H)= \sum_{\alpha_n} \hat{\rho}^{(q_1 q_2 q_3)}(\alpha_n) |\langle \alpha_n|\alpha_H\rangle|^2. \label{eq:rho123b3}
\end{align}

Inserting Eq.~(\ref{eq:rho123b3}) into Eq.~(\ref{eq:rhoHmma2}) we
can calculate the hyperon polarization and results take exactly the
same form as Eqs.~(\ref{eq:PLambda}-\ref{eq:Csigma0}) except that
all quantities are replaced by effective ones, e.g. $P_{qi}\to\bar{P}_{qi}$,
$c_{ij}^{(12)}\to\bar{c}_{ij}^{(12)}$ and so on. 
For example, the polarization of $\varLambda$ is in the form, 
\begin{equation}
P_{\varLambda}(\alpha_{\varLambda})=\bar{P}_{sz}-\frac{\bar{c}_{iiz}^{(uds)}+\bar{c}_{iz}^{(us)}\bar{P}_{di}+\bar{c}_{iz}^{(ds)}\bar{P}_{ui}}{1-\bar{c}_{ii}^{(ud)}-\bar P_{ui}\bar P_{di}}.\label{eq:bPLambda}
\end{equation}
By applying Eqs.~(\ref{eq:rho12b}) and (\ref{eq:bc123}), we can rewrite Eq.~(\ref{eq:bPLambda}) as
\begin{equation}
P_{\varLambda}(\alpha_{\varLambda})=\frac{\langle P_{sz}(1-c_{ii}^{(ud)}-P_{ui}P_{di})-c_{iiz}^{(uds)}-c_{iz}^{(us)}P_{di}-c_{iz}^{(ds)}P_{ui}\rangle_{\varLambda}}{\langle1- c_{ii}^{(ud)}-P_{ui}P_{di}\rangle_{\varLambda}},\label{eq:PLambdab2}
\end{equation}
where the averages are taken with the weight $|\langle\alpha_n|\alpha_{\varLambda}\rangle|^2$.
We see that the situation is similar to vector mesons. 
If all the genuine quark correlations $c_{ij}^{(q_{1}q_{2})}=0$ and $c_{ijk}^{(q_{1}q_{2}q_{3})}=0$,
we still have induced correlations, 
\begin{equation}
P_{\varLambda}(\alpha_{\varLambda})=\frac{\langle P_{sz}(1-P_{ui}P_{di})\rangle_{\varLambda}}{\langle1-P_{ui}P_{di}\rangle_{\varLambda}}.\label{eq:PLambdab20}
\end{equation}
Similar to the vector meson's spin alignment, we often run into $P_{\varLambda}(\alpha_{\varLambda})$
averaged over $\alpha_{\varLambda}$ in a given kinematic region. 
Then we will have an additional average over the distribution $f_\varLambda(\alpha_{\varLambda})$.
It is also obvious that we have only local quark-quark spin correlations in this case. 

Together with the results obtained in Sec.~\ref{sec:Vpol}, we see that by studying spin polarizations of one hadron 
we always study only local quark spin correlations. 

\section{Global spin correlations of hyperons \label{sec:Hcorrelation}}

The calculations can be extended in a straightforward manner to hyperon-hyperon
and hyperon-anti-hyperon spin correlations. In this section, we take
hyperon-antihyperon as an example to show the calculation and results.

For a spin-1/2 hyperon pair $H_{1}\bar{H}_{2}$, the spin correlation is usually defined in the conventional way as given by Eq.~(\ref{eq:cnndef}), i.e.,  
\begin{equation}
c_{nn}^{H_{1}\bar{H}_{2}}=\frac{f^{H_{1}\bar{H}_{2}}_{++}+f^{H_{1}\bar{H}_{2}}_{--}-f^{H_{1}\bar{H}_{2}}_{+-}-f^{H_{1}\bar{H}_{2}}_{-+}}{f^{H_{1}\bar{H}_{2}}_{++}+f^{H_{1}\bar{H}_{2}}_{--}+f^{H_{1}\bar{H}_{2}}_{+-}+f^{H_{1}\bar{H}_{2}}_{-+}},\label{eq:cHHdef}
\end{equation}
where $f^{H_{1}\bar{H}_{2}}_{m_{H_1}m_{\bar H_2}}=\langle m_{H_1}m_{\bar H_2}|\hat{\rho}^{H_{1}\bar{H}_{2}}|m_{H_1}m_{\bar H_2}\rangle $
and $m_{H_1},m_{\bar H_2}=\pm$ denoting the spin states parallel or anti-parallel to $\hat{\boldsymbol n}$-direction respectively.
We simply adopt this definition and calculate its relationship to
those quantities defined at the quark level using quark combination mechanism. 
In the calculation, the most convenient way is to rotate the Cartesian system 
so that $\hat{\boldsymbol n}$-direction is $z$-direction in the new system. 
We choose this case as an example to do the calculations and denote 
$c_{nn}^{H_{1}\bar{H}_{2}}$ in this case by $c_{zz}^{H_{1}\bar{H}_{2}}$ in the following of this paper. 
 
Now the task is to compute the spin density matrix element 
$\rho^{H_1\bar{H}_2}_{m_{H_1}m_{\bar{H}_2};m_{H_1}m_{\bar{H}_2}}
=\langle m_{H_1}m_{\bar{H}_2}|\hat{\rho}^{H_1\bar{H}_{2}}|m_{H_1}m_{\bar{H}_2}\rangle$.
Similar to spin density operators for vector mesons and hyperons given 
by Eqs.~(\ref{eq:rhoV}) and (\ref{eq:rhoH}), $\hat{\rho}^{H_{1}\bar{H}_{2}}$
is related to that of the six body system $q_{1}q_{2}q_{3}\bar{q}_{4}\bar{q}_{5}\bar{q}_{6}$ by 
\begin{equation}
\hat{\rho}^{H_{1}\bar{H}_{2}}=\hat{\mathcal{M}}\hat{\rho}^{(1\cdots6)}\hat{\mathcal{M}}^{\dagger},\label{eq:rhoH1H2}
\end{equation}
where we simply used `$1\cdots6$' to label $q_{1}q_{2}q_{3}\bar{q}_{4}\bar{q}_{5}\bar{q}_{6}$.
The matrix element of $\hat{\rho}_{H_{1}\bar{H}_{2}}$ is given by
\begin{equation}
\rho^{H_{1}\bar{H}_{2}}_{m_{H_{1}}m_{\bar{H}_{2}};m_{H_{1}}^{\prime}m_{\bar{H}_{2}}^{\prime}}=\langle j_{H_{1}}m_{H_{1}},j_{\bar{H}_{2}}m_{\bar{H}_{2}}|\hat{\mathcal{M}}\hat{\rho}^{(1\cdots6)}\hat{\mathcal{M}}^{\dagger}|j_{H_{1}}m_{H_{1}}^{\prime},j_{\bar{H}_{2}}m_{\bar{H}_{2}}^{\prime}\rangle.\label{eq:rhoH1H2m1m2}
\end{equation}
The complete expansion of $\hat{\rho}^{(1\cdots6)}$ is 
\begin{align}
\hat{\rho}^{(1\cdots6)}=& \hat{\rho}^{(1)} \otimes \hat{\rho}^{(2)} \otimes \hat{\rho}^{(3)} \otimes \hat{\rho}^{(4)}\otimes \hat{\rho}^{(5) }\otimes \hat{\rho}^{(6)} \nonumber\\     
&+\frac{1}{2^2}\text{[}c^{(12)}_{ij}\hat\sigma_{1i} \otimes \hat\sigma_{2j}\otimes \hat\rho^{(3)}\otimes \hat{\rho}^{(4)} \otimes \hat{\rho}^{(5)}\otimes \hat{\rho}^{(6)}+14~\text{exchange terms}\text{ ]} \nonumber\\  
&+\frac{1}{2^3}\text{[}c^{(123)}_{ijk}\hat\sigma_{1i}\otimes\hat\sigma_{2j}\otimes\hat\sigma_{3k}\otimes \hat{\rho}^{(4)} \otimes \hat{\rho}^{(5)} \otimes \hat{\rho}^{(6)}+19~\text{exchange terms}\text{ ]} \nonumber\\     
&+\frac{1}{2^4}\text{[}c^{(1234)}_{ijkl}\hat\sigma_{1i}\otimes\hat\sigma_{2j}\otimes\hat\sigma_{3k}\otimes \hat\sigma_{4l} \otimes \hat{\rho}^{(5)} \otimes \hat{\rho}^{(6)} +14~\text{exchange terms}\text{ ]}  \nonumber\\       
&+\frac{1}{2^5}\text{[}c^{(12345)}_{ijklm}\hat\sigma_{1i} \otimes \hat\sigma_{2j} \otimes \hat\sigma_{3k} \otimes \hat\sigma_{4l}\otimes \hat\sigma_{5m}\otimes \hat{\rho}^{(6)}+5~\text{exchange terms}  \text{ ]}     \nonumber \\   
&+\frac{1}{2^6}c^{(123456)}_{ijklmn}\hat\sigma_{1i}\otimes\hat\sigma_{2j}\otimes\hat\sigma_{3k}\otimes\hat\sigma_{4l}\otimes\hat\sigma_{5m}\otimes\hat\sigma_{6n}.     \label{eq:rho6q}
\end{align}

In the following of this section, we take $\varLambda\bar{\varLambda}$
as an example to show the calculation of the hyperon-anti-hyperon spin correlation. 
For simplicity, we only consider two-particle spin correlations and set all others with more than two particles as zero. 
As before, we consider two cases, the one with only spin degree of freedom and the one with other degrees
of freedom denoted by $\alpha$. 
The calculations can be extended to other hyperons and/or include spin correlations of more than two particles in a straightforward way.

\subsection{With spin degree of freedom}

As in previous sections, we insert the completeness identity $\sum_{m_n}|m_n\rangle\langle m_n|=1$
into Eq.~(\ref{eq:rhoH1H2m1m2}), and obtain 
\begin{align}
 & \rho^{H_{1}\bar{H}_{2}}_{m_{H_{1}}m_{\bar{H}_{2}};m_{H_{1}}^{\prime}m_{\bar{H}_{2}}^{\prime}}
= \sum_{m_n,m_n^{\prime}}\langle m_{H_{1}}m_{\bar{H}_{2}}|\hat{\mathcal{M}}|m_n\rangle\langle m_n|\hat{\rho}^{(1\cdots6)}|m_n^{\prime}\rangle\langle m_n^{\prime}|\hat{\mathcal{M}}^{\dagger}|m_{H_{1}}^{\prime}m_{\bar{H}_{2}}^{\prime}\rangle,\label{eq:rhoH1H2m1m2a}
\end{align}
where we have suppressed $j_{H_{1}}=j_{\bar{H}_{2}}=1/2$ for $H_{1}=\varLambda$ and $\bar{H}_{2}=\bar{\varLambda}$. 
The transition matrix element can be written as 
\begin{align}
 \langle m_{H_{1}}m_{\bar{H}_{2}}|\hat{\mathcal{M}}|m_n\rangle 
= &\langle m_{H_{1}} m_{\bar{H}_{2}}|\hat{\mathcal{M}}|m_{H_{1}}m_{\bar{H}_{2}}\rangle\langle m_{H_{1}}m_{\bar{H}_{2}}|m_n\rangle\nonumber \\
= & \langle m_{H_{1}}|\hat{\mathcal{M}}_{H}|m_{H_{1}}\rangle\langle m_{\bar{H}_{2}}|\hat{\mathcal{M}}_{\bar{H}}|m_{\bar{H}_{2}}\rangle\langle m_{H_{1}}m_{\bar{H}_{2}}|m_n\rangle,\label{eq:MHHfac}
\end{align}
where we have assumed a factorization form for $\langle m_{H_{1}}m_{\bar{H}_{2}}|\hat{\mathcal{M}}|m_{H_{1}}m_{\bar{H}_{2}}\rangle$
with $\hat{\mathcal{M}}=\hat{\mathcal{M}}_{H}\hat{\mathcal{M}}_{\bar{H}}$, so that
the transition matrix contributes only to the normalization constant
and has no effect on the spin part. Then we obtain 
\begin{align}
 & \rho^{H_{1}\bar{H}_{2}}_{m_{H_{1}}m_{\bar{H}_{2}};m_{H_{1}}^{\prime}m_{\bar{H}_{2}}^{\prime}}
= N_{H_{1}\bar{H}_{2}}\sum_{m_n,m_n^{\prime}}\langle m_{H_{1}}m_{\bar{H}_{2}}|m_n\rangle\langle m_n|\hat{\rho}^{(1\cdots6)}|m_n^{\prime}\rangle\langle m_n^{\prime}|m_{H_{1}}^{\prime}m_{\bar{H}_{2}}^{\prime}\rangle.\label{eq:rhoH1H2m1m2b}
\end{align}
We note that in production processes of $H_{1}$ and $\bar{H}_{2}$,
$q_{1}+q_{2}+q_{3}\to H_{1}$ and $\bar{q}_{4}+\bar{q}_{5}+\bar{q}_{6}\to\bar{H}_{2}$,
the Clebsch-Gordan coefficient $\langle m_{H_{1}}m_{\bar{H}_{2}}|m_n\rangle$
is just the product of $\langle m_{H_{1}}|m_1m_2m_3\rangle$ and $\langle m_{\bar{H}_{2}}|m_4m_5m_6\rangle$.

When all two-particle spin correlations are considered, the result for the spin correlation of $\varLambda\bar{\varLambda}$ is 
\begin{align}
c_{zz}^{\varLambda\bar{\varLambda}}= & P_{sz}P_{\bar{s}z}+\frac{1}{C_{\varLambda\bar{\varLambda}}}\left\{ c_{zz}^{(s\bar{s})}(1-P_{ui}P_{di})(1-P_{\bar{u}i}P_{\bar{d}i})\right.\nonumber \\
 & -P_{sz}\left[(c_{iz}^{(d\bar{s})}P_{ui}+c_{iz}^{(u\bar{s})}P_{di})(1-P_{\bar{u}i}P_{\bar{d}i})+(c_{iz}^{(\bar{d}\bar{s})}P_{\bar{u}i}+c_{iz}^{(\bar{u}\bar{s})}P_{\bar{d}i})(1-P_{ui}P_{di})\right]\nonumber \\
 & \left.-P_{\bar{s}z}\left[(c_{iz}^{(ds)}P_{ui}+c_{iz}^{(us)}P_{di})(1-P_{\bar{u}i}P_{\bar{d}i})+(c_{iz}^{(\bar{d}s)}P_{\bar{u}i}+c_{iz}^{(\bar{u}s)}P_{\bar{d}i})(1-P_{ui}P_{di})\right]\right\} ,\label{eq:cLamLamB}
\end{align}
where the normalization constant $C_{\varLambda\bar{\varLambda}}$ is given by 
\begin{align}
C_{\varLambda\bar{\varLambda}}= 
C_\varLambda C_{\bar\varLambda}-c_{ii}^{(ud)} c_{jj}^{(\bar{u}\bar{d})}
 +c_{ij}^{(u\bar{u})}P_{di}P_{\bar{d}j}+c_{ij}^{(d\bar{d})}P_{ui}P_{\bar{u}j}+c_{ij}^{(d\bar{u})}P_{ui}P_{\bar{d}j}+c_{ij}^{(u\bar{d})}P_{di}P_{\bar{u}j},\label{eq:CLamLamB}
\end{align}
where $C_\varLambda$ is given by Eq.~(\ref{eq:CLambda}) and $C_{\bar\varLambda}$ is obtained from $C_\varLambda$ 
with the replacement of everything by that of the corresponding anti-quark.  

We compare the results given by Eqs.~(\ref{eq:cLamLamB}-\ref{eq:CLamLamB}) with those given by 
Eqs.~(\ref{eq:PLambda}-\ref{eq:CLambda}) for $\varLambda$-polarization. 
We note that we need to put all spin correlations of more than two quarks and/or anti-quarks 
and products of two particle spin correlations as zero since we consider only two particle spin correlations. 
In this way, we obtain, 
\begin{equation}
c_{zz}^{\varLambda\bar{\varLambda}}\approx P_{\varLambda z}P_{\bar{\varLambda}z}+c^{(s\bar s)}_{zz} 
-\frac{P_{sz} }{C_{\varLambda}} \bigl[c_{iz}^{(d\bar{s})}P_{ui}+c_{iz}^{(u\bar{s})}P_{di}\bigr]
-\frac{P_{\bar{s}z}}{C_{\bar\varLambda}} \bigl[c_{iz}^{(s\bar d)}P_{\bar ui}+c_{iz}^{(s\bar u)}P_{\bar di}\bigr]. \label{eq:cLamLamBa}
\end{equation}
From Eq.~(\ref{eq:cLamLamBa}), we see clearly that the spin correlation of $\varLambda$ and $\bar\varLambda$ 
comes from those of quarks and anti-quarks.
We also see clearly that $c_{zz}^{\varLambda\bar\varLambda}= P_{\varLambda z}P_{\bar{\varLambda}z}$ 
if only quark-quark and anti-quark-anti-quark spin correlations are considered.

\subsection{With other degrees of freedom}

It is clear that in this case the six-quark (anti-quark) spin density
matrix $\hat{\rho}^{(1\cdots6)}(\alpha_n)$ takes the same
form as $\hat{\rho}^{(1\cdots6)}$ in (\ref{eq:rho6q}) except that
all $\hat{\rho}^{(n)}$ ($n=1,\cdots,6$) depends on $\alpha_n$
and that all correlation coefficients $c_{i_1\cdots i_n}^{(1\cdots n)}$
with $n\leq6$ depend on $\alpha_n$. 
The spin density
matrix for $H_{1}\bar{H}_{2}$ now becomes $\rho^{H_{1}\bar{H}_{2}}_{m_{H_{1}}m_{\bar{H}_{2}};m_{H_{1}}^{\prime}m_{\bar{H}_{2}}^{\prime}}(\alpha_{H_{1}},\alpha_{\bar{H}_{2}})$
that depends on $\alpha_{H_{1}}$ and $\alpha_{\bar{H}_{2}}$. Then
we obtain a similar formula for $\rho^{H_{1}\bar{H}_{2}}_{m_{H_{1}}m_{\bar{H}_{2}};m_{H_{1}}^{\prime}m_{\bar{H}_{2}}^{\prime}} (\alpha_{H_{1}},\alpha_{\bar{H}_{2}})$ to Eq.~(\ref{eq:rhoH1H2m1m2b}). 
By assuming a factorization condition
similar to Eq. (\ref{eq:MHHfac}) for $H_{1}$ and $\bar{H}_{2}$
and that for the spin and $\alpha$ parts of wave functions, we obtain
\begin{align}
 \rho^{H_{1}\bar{H}_{2}}_{m_{H_{1}}m_{\bar{H}_{2}};m_{H_{1}}^{\prime}m_{\bar{H}_{2}}^{\prime}} & (\alpha_{H_{1}},\alpha_{\bar{H}_{2}})
=N_{H_{1}\bar{H}_{2}}(\alpha_{H_{1}},\alpha_{\bar{H}_{2}})\nonumber\\
&\times\sum_{m_n,m_n^{\prime}}\langle m_{H_{1}}m_{\bar{H}_{2}}|m_n\rangle 
 \langle m_n|\hat{\bar{\rho}}^{(1\cdots6)}(\alpha_{H_{1}},\alpha_{\bar{H}_{2}})|m_n^{\prime}\rangle\langle m_n^{\prime}|m_{H_{1}}^{\prime}m_{\bar{H}_{2}}^{\prime}\rangle,\label{eq:rhoH1H2m1m2c}
\end{align}
where the effective density matrix is given by 
\begin{align}
\hat{\bar{\rho}}^{(1\cdots6)}(\alpha_{H_{1}},\alpha_{\bar{H}_{2}})= & \sum_{\alpha_n}\sum_{\alpha_m}\hat{\rho}^{(1\cdots6)}(\alpha_n,\alpha_m)|\langle\alpha_n|\alpha_{H_{1}}\rangle|^2 |\langle\alpha_m|\alpha_{\bar{H}_{2}}\rangle|^2.\label{eq:arho16}
\end{align}
We see the difference between Eq.~(\ref{eq:rhoH1H2m1m2b}) for the
case with only the spin degree of freedom and Eq. (\ref{eq:rhoH1H2m1m2c})
is the replacement $\hat{\rho}^{(1\cdots6)}\rightarrow\hat{\bar{\rho}}^{(1\cdots6)}(\alpha_{H_{1}},\alpha_{\bar{H}_{2}})$.

We emphasize the average in Eq.~(\ref{eq:arho16}) can be carried
out inside $H_{1}$ and $\bar{H}_{2}$ for quarks and antiquarks successively.
This is different from the case for vector meson discussed in Sec.
\ref{sec:rhoValpha} where the average inside $V$ is carried out
for the quark and antiquark simultaneously. 
More precisely, we have
\begin{align}
\bar{P}_{q_{l}}(\alpha_{H_{1}})= & \sum_{\alpha_n}P_{q_{l}}(\alpha_{q_l})|\langle\alpha_n|\alpha_{H_{1}}\rangle|^2=\langle P_{q_{l}}(\alpha_{q_l})\rangle_{H_{1}},\label{eq:PqiH1}\\
\bar{P}_{\bar{q}_{l}}(\alpha_{\bar{H}_{2}})= & \sum_{\alpha_m}P_{\bar{q}_{l}}(\alpha_{\bar q_l})|\langle\alpha_m|\alpha_{\bar{H}_{2}}\rangle|^2 =\langle P_{\bar{q}_{l}}(\alpha_{\bar q_l})\rangle_{\bar{H}_{2}}.\label{eq:PqiH2}
\end{align}
We obtain two-particle spin correlations as 
\begin{align}
\bar{c}_{ij}^{(q_{1}q_{2})}(\alpha_{H_{1}},\alpha_{\bar{H}_{2}})= & \sum_{\alpha_n,\alpha_m}\left[c_{ij}^{(q_{1}q_{2})}(\alpha_{q_{1}},\alpha_{q_{2}})+P_{q_{1}i}(\alpha_{q_{1}})P_{q_{2}j}(\alpha_{q_{2}})\right] \left|\langle\alpha_n|\alpha_{H_{1}}\rangle\right|^{2}\left|\langle\alpha_m|\alpha_{\bar{H}_{2}}\rangle\right|^{2}-\bar{P}_{q_{1}i}(\alpha_{H_{1}})\bar{P}_{q_{2}j}(\alpha_{H_{1}}),\nonumber \\
= & \sum_{\alpha_n}\left[c_{ij}^{(q_{1}q_{2})}(\alpha_{q_{1}},\alpha_{q_{2}})+P_{q_{1}i}(\alpha_{q_{1}})P_{q_{2}j}(\alpha_{q_{2}})\right]\left|\langle\alpha_n|\alpha_{H_{1}}\rangle\right|^{2}-\bar{P}_{q_{1}i}(\alpha_{H_{1}})\bar{P}_{q_{2}j}(\alpha_{H_{1}}),\label{eq:cijq1q2}\\
\bar{c}_{ij}^{(\bar{q}_{1}\bar{q}_{2})}(\alpha_{H_{1}},\alpha_{\bar{H}_{2}})= & \sum_{\alpha_n,\alpha_m}\left[c_{ij}^{(\bar{q}_{1}\bar{q}_{2})}(\alpha_{\bar{q}_{1}},\alpha_{\bar{q}_{2}})+P_{\bar{q}_{1}i}(\alpha_{\bar{q}_{1}})P_{\bar{q}_{2}j}(\alpha_{\bar{q}_{2}})\right]
\left|\langle\alpha_n|\alpha_{H_{1}}\rangle\right|^{2}\left|\langle\alpha_m|\alpha_{\bar{H}_{2}}\rangle\right|^{2}-\bar{P}_{\bar{q}_{1}i}(\alpha_{\bar{H}_{2}})\bar{P}_{\bar{q}_{2}j}(\alpha_{\bar{H}_{2}}),\nonumber \\
= & \sum_{\alpha_m}\left[c_{ij}^{(\bar{q}_{1}\bar{q}_{2})}(\alpha_{\bar{q}_{1}},\alpha_{\bar{q}_{2}})+P_{\bar{q}_{1}i}(\alpha_{\bar{q}_{1}})P_{\bar{q}_{2}j}(\alpha_{\bar{q}_{2}})\right]\left|\langle\alpha_m|\alpha_{\bar{H}_{2}}\rangle\right|^{2} 
-\bar{P}_{\bar{q}_{1}i}(\alpha_{\bar{H}_{2}})\bar{P}_{\bar{q}_{2}j}(\alpha_{\bar{H}_{2}}),\label{eq:cijBq1Bq2}\\
\bar{c}_{ij}^{(q_{1}\bar{q}_{2})}(\alpha_{H_{1}},\alpha_{\bar{H}_{2}})= & \sum_{\alpha_n,\alpha_m}\left[c_{ij}^{(q_{1}\bar{q}_{2})}(\alpha_{q_{1}},\alpha_{\bar{q}_{2}})+P_{q_{1}i}(\alpha_{q_{1}})P_{\bar{q}_{2}j}(\alpha_{\bar{q}_{2}})\right]
\left|\langle\alpha_n|\alpha_{H_{1}}\rangle\right|^{2}\left|\langle\alpha_m|\alpha_{\bar{H}_{2}}\rangle\right|^{2}-\bar{P}_{q_{1}i}(\alpha_{H_{1}})\bar{P}_{\bar{q}_{2}j}(\alpha_{\bar{H}_{2}})\nonumber \\
= & \sum_{\alpha_n,\alpha_m}c_{ij}^{(q_{1}\bar{q}_{2})}(\alpha_{q_{1}},\alpha_{\bar{q}_{2}})\left|\langle\alpha_n|\alpha_{H_{1}}\rangle\right|^{2}\left|\langle\alpha_m|\alpha_{\bar{H}_{2}}\rangle\right|^{2}.\label{eq:cijq1Bq2}
\end{align}
We see that $\bar{c}_{ij}^{(q_{1}q_{2})}$ is independent of $\alpha_{\bar{H}_{2}}$
and $\bar{c}_{ij}^{(\bar{q}_{1}\bar{q}_{2})}$ is independent of $\alpha_{H_{1}}$,
while $\bar{c}_{ij}^{(q_{1}\bar{q}_{2})}$ just reduces to 
\begin{equation}
\bar{c}_{ij}^{(q_{1}\bar{q}_{2})}(\alpha_{H_{1}},\alpha_{\bar{H}_{2}})=\langle c_{ij}^{(q_{1}\bar{q}_{2})}(\alpha_{q_{1}},\alpha_{\bar{q}_{2}})\rangle_{H_{1}\bar{H}_{2}},\label{eq:cijq1Bq2a}
\end{equation}
because 
\begin{equation}
\langle P_{q_{1}i}(\alpha_{q_{1}})P_{\bar{q}_{2}j}(\alpha_{\bar{q}_{2}})\rangle_{H_{1}\bar{H}_{2}}=\langle P_{q_{1}i}(\alpha_{q_{1}})\rangle_{H_{1}}\langle P_{\bar{q}_{2}j}(\alpha_{\bar{q}_{2}})\rangle_{\bar{H}_{2}}.
\end{equation}
Here we neglect the overlap of $H_{1}$ and $\bar{H}_{2}$ in $\alpha$-space.
In this case, we have no contributions from the induced spin correlation
between the quark and anti-quark, i.e., $\bar{c}_{ij}^{(q_{1}\bar{q}_{2};0)}(\alpha_{H_{1}},\alpha_{\bar{H}_{2}})=0$.
Also, because $\alpha_{q_{i}}$ is inside $H_{1}$ while $\alpha_{\bar{q}_{j}}$
is inside $\bar{H}_{2}$, we do not have contributions from local spin
correlations between quarks and antiquarks. 
This can be seen more clearly if we assume all genuine two-particle correlations vanish.
In this case, the average of each term in Eqs.~(\ref{eq:arho16})
can be separated into a product two factors, one is inside $H_{1}$
for quarks and the other is inside $\bar{H}_{2}$ for anti-quarks.
This shows explicitly that we have contributions from local quark-quark
and antiquark-antiquark correlations but no contribution from local quark-antiquark correlations. 
Hence in spin correlations between the hyperon and anti-hyperon, there is no contribution from local correlations
between the quark and antiquark. 

Now we compute the spin correlation of $\varLambda\bar{\varLambda}$ with above formula. 
With only two-particle spin correlations, the result is just those given by Eqs.~(\ref{eq:cLamLamB}-\ref{eq:cLamLamBa}) 
with the replacement of all quantities by the corresponding effective ones, i.e. 
\begin{equation}
c_{zz}^{\varLambda\bar{\varLambda}}(\alpha_\varLambda,\alpha_{\bar\varLambda})
\approx P_{\varLambda z}(\alpha_\varLambda)P_{\bar{\varLambda}z}(\alpha_{\bar\varLambda})+\bar c^{(s\bar s)}_{zz} 
-\frac{\bar P_{sz} }{\bar C_{\varLambda}} \bigl[\bar c_{iz}^{(d\bar{s})}\bar P_{ui}+\bar c_{iz}^{(u\bar{s})}\bar P_{di}\bigr]
-\frac{\bar P_{\bar{s}z}}{\bar C_{\bar\varLambda}} \bigl[\bar c_{iz}^{(s\bar d)}\bar P_{\bar ui}+\bar c_{iz}^{(s\bar u)}\bar P_{\bar di}\bigr]. \label{eq:cLamLamBb}
\end{equation}
We emphasize in particular that all quantities for quarks and/or anti-quarks on the r.h.s. of Eq.~(\ref{eq:cLamLamBb}) are effective ones 
and are functions of $\alpha_\varLambda$ and/or $\alpha_{\bar\varLambda}$.    
More precisely, $\bar P_{qi}$ and $\bar P_{\bar qi}$ are functions of $\alpha_\varLambda$ and  $\alpha_{\bar\varLambda}$ respectively,  
while $\bar c_{ij}^{(q_1\bar q_2)}$ is a function of $(\alpha_\varLambda,\alpha_{\bar\varLambda})$.
These results are valid in the case that we neglect the overlap of the wave function of $\varLambda$ with that of $\bar\varLambda$.  

We now discuss a simple case where both quark polarizations and quark spin correlations are small so that we can neglect 
the last two terms in Eq.~(\ref{eq:cLamLamBb}) compared with the first two. 
In this case, when further averaged over $\alpha_\varLambda$ and $\alpha_{\bar\varLambda}$ in a given kinematic region, we obtain
\begin{equation}
\langle c_{zz}^{\varLambda\bar{\varLambda}}\rangle \approx 
\langle P_{\varLambda z}(\alpha_\varLambda) P_{\bar{\varLambda}z}(\alpha_{\bar\varLambda})\rangle +\langle\bar c^{(s\bar s)}_{zz}\rangle
\approx 
\langle P_{\varLambda z}\rangle \langle P_{\bar{\varLambda}z}\rangle
+\langle c^{s\bar s;0}_{zz}\rangle+\langle\bar c^{(s\bar s)}_{zz}\rangle, \label{eq:cLamLamBc}
\end{equation}
where $\langle \bar c_{zz}^{(s\bar{s};0)}\rangle
=\langle\bar{P}_{sz}(\alpha_{\varLambda})\bar{P}_{\bar{s}z}(\alpha_{\bar{\varLambda}})\rangle - 
\langle\bar{P}_{sz}(\alpha_{\varLambda})\rangle\langle\bar{P}_{\bar{s}z}(\alpha_{\bar\varLambda})\rangle$. 
Compared to those inside a hadron, we call this long range correlation. 
We see that $\langle \bar c_{zz}^{(s\bar{s};0)}\rangle$ is the contribution from the induced spin correlation  
while $\langle\bar c^{(s\bar s)}_{zz}\rangle$ is from the genuine quark spin correlation of $s\bar s$. 

Since we do not consider the overlap of the wave functions of the hyperon and that of the anti-hyperon, 
the results obtained above can be extended directly to hyperon-hyperon spin correlations. 
In particular, those given by Eqs.~(\ref{eq:cLamLamBb}-\ref{eq:cLamLamBc}) 
can be extended to $\varLambda\varLambda$ spin correlations if we neglect the overlap of the wave function of the two $\varLambda$'s. 
In this case, we need only to replace $\bar P_{q}(\alpha_{\varLambda})$ and $\bar P_{\bar q}(\alpha_{\bar\varLambda})$ 
by $\bar P_{q}(\alpha_{\varLambda_1})$ and $\bar P_{q}(\alpha_{\varLambda_2})$ respectively 
in order to obtain $c_{zz}^{\varLambda\varLambda}(\alpha_{\varLambda_1},\alpha_{\varLambda_2})$. 
\begin{align}
c_{zz}^{\varLambda{\varLambda}}(\alpha_{\varLambda_1},\alpha_{\varLambda_2}) \approx 
& P_{\varLambda z}(\alpha_{\varLambda_1})P_{\varLambda z}(\alpha_{\varLambda_2})+\bar c^{(ss)}_{zz} (\alpha_{\varLambda_1},\alpha_{\varLambda_2}) \nonumber\\ 
&-\frac{\bar P_{sz}(\alpha_{\varLambda_1}) }{\bar C_{\varLambda}(\alpha_{\varLambda_1})} 
\bigl[\bar c_{iz}^{(ds)}(\alpha_{\varLambda_1},\alpha_{\varLambda_2})\bar P_{ui}(\alpha_{\varLambda_1})+\bar c_{iz}^{(us)}(\alpha_{\varLambda_1},\alpha_{\varLambda_2})\bar P_{di}(\alpha_{\varLambda_1})\bigr] \nonumber\\
&-\frac{\bar P_{sz}(\alpha_{\varLambda_2})}{\bar C_{\varLambda}(\alpha_{\varLambda_2})} \bigl[\bar c_{iz}^{(sd)}(\alpha_{\varLambda_1},\alpha_{\varLambda_2})\bar P_{ui}(\alpha_{\varLambda_2})+\bar c_{iz}^{(su)}(\alpha_{\varLambda_1},\alpha_{\varLambda_2})\bar P_{di}(\alpha_{\varLambda_2})\bigr]. \label{eq:cLamLam}
\end{align}
In the simple case considered above for  $\langle c_{zz}^{\varLambda\bar{\varLambda}}\rangle$ we obtain similar result 
for $c_{zz}^{\varLambda{\varLambda}}$ as 
\begin{equation}
\langle c_{zz}^{\varLambda{\varLambda}}\rangle \approx 
\langle P_{\varLambda z}\rangle ^2 
+\langle \bar c^{(ss;0)}_{zz}\rangle+\langle\bar c^{(ss)}_{zz}\rangle. \label{eq:cLamLama}
\end{equation}
We see that in this case the spin correlation between $\varLambda$'s measures the spin correlation between $s$-quarks.  

To compare with the results obtained in Sec.~\ref{sec:Vpol}, we see clearly that by studying the spin alignment of $\phi$ meson,  
we study the spin correlations between $s$ and $\bar s$ inside the vector meson.    
In contrast, by studying $\varLambda\bar\varLambda$ and $\varLambda\varLambda$ spin correlations, 
we study the spin correlations between $ss$ or $s\bar s$ in the whole QGP system~\cite{Pang:2016igs,Sheng:2022wsy,Sheng:2022ffb}. 
The former is in general short ranged while the latter includes long range contributions. 
The strength of such correlations is determined by the dynamics of the system and is an important direction for future study.  

\end{widetext}

\section{Numerical estimates \label{sec:numerical}}

The global quark spin polarizations and correlations are determined by the QCD dynamics in heavy-ion collisions  
and can be calculated using QCD-based theoretical models. 
Having the relationships between measurable quantities at the hadron level and global spin
properties at the quark level, we can also extract them from data available and make predictions for other measurable quantities.
The available data are however still far from enough to make high precision predictions. 
In this section, we just present a rough estimate based on the data 
available~\citep{STAR:2017ckg,STAR:2018gyt,STAR:2020xbm,STAR:2021beb,ALICE:2019onw,HADES:2022enx,STAR:2022fan}. 

We use Eqs.~(\ref{eq:PLambdab20}, \ref{eq:rhoV00a2}, \ref{eq:cLamLamBc}) and take approximately 
\begin{align}
 & \langle P_{\varLambda} \rangle \sim \langle P_{s} \rangle, \label{eq:appPLambda}\\
 & \langle \rho_{00}^{\phi}\rangle \sim\frac{1-\bar{c}_{zz;\phi}^{(s\bar{s})}-\langle P_{s}\rangle^2}  {3+\bar{c}_{zz;\phi}^{(s\bar{s})} +\langle P_{s}\rangle^2},\label{eq:apprho00phi}\\
 & \langle c_{zz}^{\varLambda\bar{\varLambda}}\rangle \sim\bar{c}_{zz;\varLambda\bar\varLambda}^{(s\bar s)}+\langle P_s\rangle^2, \label{eq:appCLamLamB}
\end{align}
where all quark spin correlations are effective ones and are in general sums of genuine and induced contributions. 
We use these equations to extract $\langle P_{s}\rangle$ and $\bar{c}_{zz;\phi}^{(s\bar{s})}$ 
from data available on $\langle P_\varLambda\rangle$ and $\langle\rho_{00}^{\phi}\rangle$~\citep{STAR:2017ckg,STAR:2018gyt,STAR:2020xbm,STAR:2021beb,ALICE:2019onw,HADES:2022enx,STAR:2022fan}, 
and make estimates of $\langle c_{zz}^{\varLambda\bar{\varLambda}}\rangle$.  

\begin{figure}
\centering	
\includegraphics[scale=0.55]{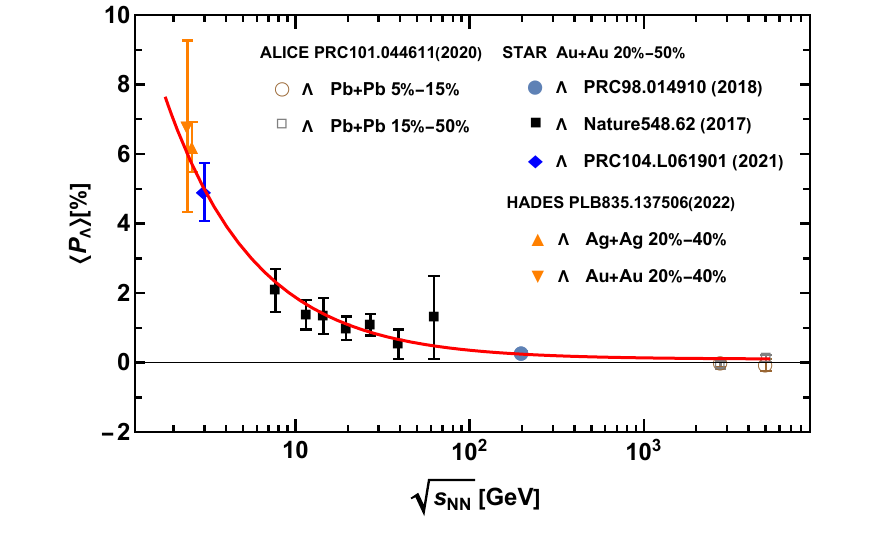}\centering	
\caption{Fit to the global $\varLambda$ polarization as a function of energy $\sqrt{s_{NN}}$. 
The data are taken from Refs.~\citep{STAR:2017ckg,STAR:2018gyt,STAR:2020xbm,STAR:2021beb,ALICE:2019onw,HADES:2022enx}.
\label{fig:P_qz}}
\end{figure}

\begin{figure}
\centering	
\includegraphics[scale=0.5]{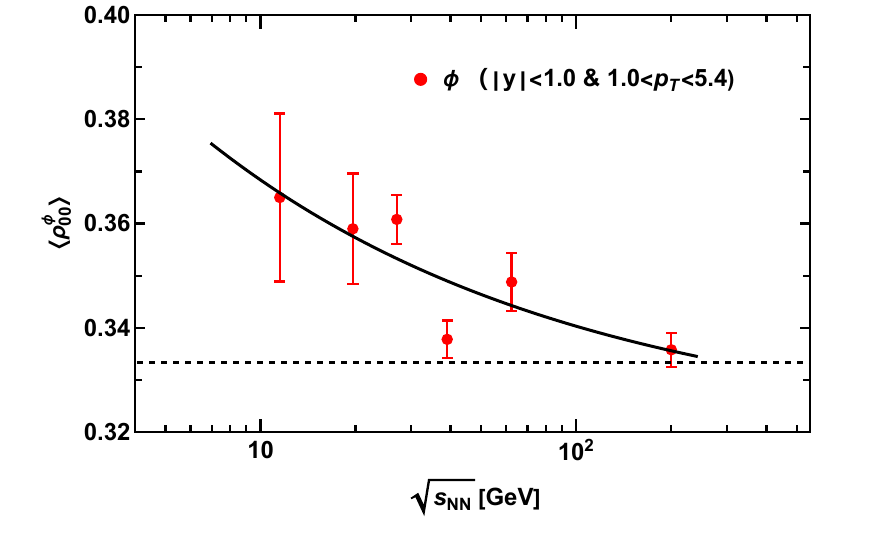}\centering	
\caption{Fit to $\rho_{00}^{\phi}$ as a function of energy $\sqrt{s_{NN}}$.
The data are taken from Ref.~\citep{STAR:2022fan}. 
\label{fig:rho_00}}
\end{figure}

\begin{figure}
\centering	
\includegraphics[scale=0.32]{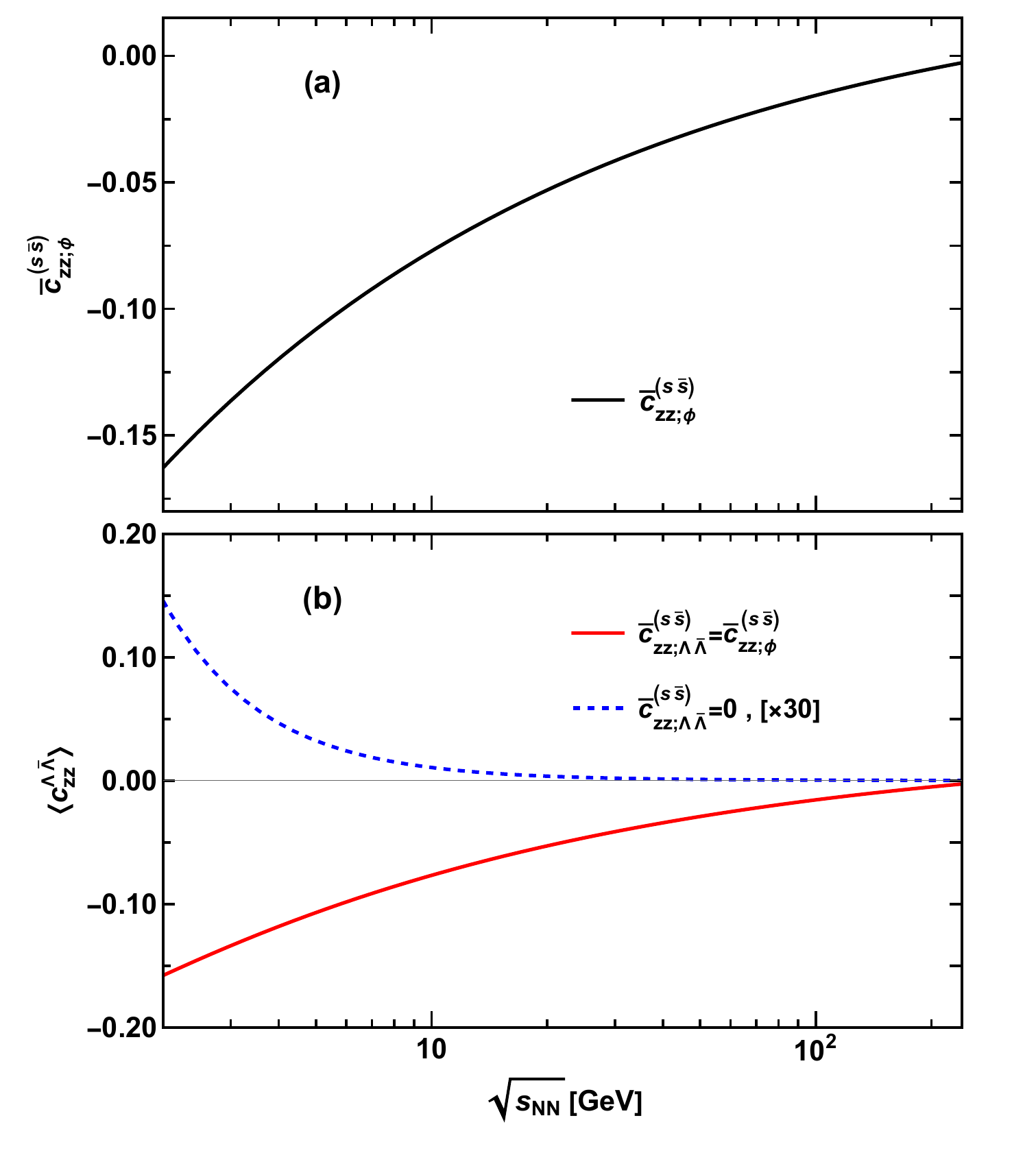}\centering	
\caption{(a) The effective global spin correlation $\bar{c}_{zz}^{(s\bar{s})}$ between $s$ and $\bar{s}$ as a function of energy $\sqrt{s_{NN}}$
obtained by fitting the data~\citep{STAR:2017ckg,STAR:2018gyt,STAR:2020xbm,STAR:2021beb,ALICE:2019onw,HADES:2022enx,STAR:2022fan}
using Eqs.~(\ref{eq:appPLambda}) and (\ref{eq:apprho00phi}) described in the text. 
(b) Estimates of $\langle {c}_{zz}^{\varLambda\bar{\varLambda}}\rangle$ 
as functions of $\sqrt{s_{NN}}$ in the two extreme cases described in the text. 
\label{fig:czz}}
\end{figure}

We take the following forms of $\langle P_{s}\rangle$ and $\bar{c}_{zz;\phi}^{(s\bar{s})}$ as functions of $\sqrt{s_{NN}}$, 
\begin{align}
 & \langle P_{s}\rangle =a s_{NN}^{-b}+c, \label{eq:pqnumberical}\\
 & \bar{c}_{zz;\phi}^{(s\bar{s})}=e s_{NN}^{-f}+d.\label{eq:cssnumerical}
\end{align}
By taking $a=0.123$, $b=0.42$, $c=0.002$, $d=0.032$, $e=-0.25$, and $f=0.18$, we obtain the fits
to $\langle P_{\varLambda}\rangle$ and $\langle \rho_{00}^{\phi}\rangle$ as shown in Figs.~\ref{fig:P_qz} and \ref{fig:rho_00} respectively. 
The obtained $\bar{c}_{zz;\phi}^{(s\bar{s})}$ as a function of $\sqrt{s_{NN}}$ is shown in Fig.~\ref{fig:czz}(a).


We take two extreme examples, i.e., $\bar{c}_{zz;\varLambda\bar\varLambda}^{(s\bar s)}=\bar{c}_{zz;\phi}^{(s\bar{s})}$ or $\bar{c}_{zz;\varLambda\bar\varLambda}^{(s\bar s)}=0$ 
and draw the results for $\langle c_{zz}^{\varLambda\bar{\varLambda}}\rangle$ as functions of $\sqrt{s_{NN}}$ shown in Fig.~\ref{fig:czz}(b). 
We see that the results in these two extreme cases are quite different from each other and they can be tested by future experiments. 

\section{Summary and outlook \label{sec:summary}}

The STAR measurements of the global spin alignment of vector mesons $\rho_{00}^{\phi}$~\citep{STAR:2022fan} indicate that there are
strong global quark-antiquark spin correlations in relativistic heavy-ion collisions. 
It opens a new window to study properties of QGP and reaction mechanisms of relativistic heavy ion collisions. 
We propose a systematic way of describing quark and/or antiquark spin correlations in the QGP. 
We show that effective quark spin correlations contain contributions
from genuine spin correlations from dynamics and induced spin correlations due to average over other degrees of freedom. 
We derive the relationships between these spin correlations at the quark level 
and those for hyperons and vector mesons that are measurable in experiments. 
We show in particular that the vector meson's spin density matrix elements, either diagonal or off-diagonal, 
are sensitive to local spin correlations between the quark and antiquark,
while hyperon-anti-hyperon spin correlations are sensitive to long rang quark spin correlations. 
We present a rough estimate of spin correlations based on available 
data~\citep{STAR:2017ckg,STAR:2018gyt,STAR:2020xbm,STAR:2021beb,ALICE:2019onw,HADES:2022enx,STAR:2022fan} to guide future measurements.

We point out that genuine spin correlations have never been considered in most theoretical studies of spin phenomena in heavy-ion collisions
to our knowledge~\citep{Wang:2023fvy,Chen:2023hnb,talks,Li-Juan:2023bws,Xin-Li:2023gwh,Yang:2017sdk,Sheng:2019kmk,Sheng:2020ghv,Sheng:2022wsy,Sheng:2022ffb,Xia:2020tyd,Wei:2023pdf,Kumar:2023ghs,DeMoura:2023jzz,Fu:2023qht,Sheng:2023urn,Fang:2023bbw,Dong:2023cng,Kumar:2023ojl}.  
The global vector meson's spin alignment in previous studies comes only from induced quark correlations. 
We note that genuine spin correlations exist in general for quarks and/or anti-quarks produced in elementary high energy processes 
such as $e^+e^-\to q\bar q$~\cite{Augustin:1978wf,Chen:2016iey} and have been discussed in connection with dihadron spin correlations 
in $e^+e^-$, $lp$ or $pp$ collisions~\cite{Chen:1994ar,Zhang:2023ugf,Gong:2021bcp,Li:2023qgj}.   
The STAR data~\citep{STAR:2022fan} suggest a strong global quark-antiquark spin correlation, 
so the study of global genuine quark spin correlations in heavy-ion collisions at the dynamical level can be an important direction in the future.

When another degree of freedom characterized by $\alpha$ is considered,
we assume that the spin and $\alpha$ part of the wave function are factorizable. 
This is general true in the non-relativistic case. 
However, in the relativistic case, spin and other degree of freedom such as
momentum are usually coupled in an intrinsic way so that such a factorization is impossible. 
In such cases, the calculation is more complicated but can be done, which we reserve for a future study.

\begin{acknowledgments}
This work was supported in part by the National Natural Science Foundation
of China (approval Nos. 12375075,  12135011, 12321005, 11890713,  11890710) 
and Shandong Province Natural Science Foundation.
\end{acknowledgments}



\end{document}